# Post-pandemic social contacts in Italy: implications for social distancing measures on in-person school and work attendance

Lorenzo Lucchini[1,*], Valentina Marziano[2,*], Filippo Trentini[1], Chiara Chiavenna[1], Elena D'Agnese[1,3,4], Vittoria Offeddu[1], Mattia Manica[2], Piero Poletti[2], Duilio Balsamo[1], Giorgio Guzzetta[2], Marco Ajelli[5,#], Alessia Melegaro[1,#], Stefano Merler[2,#]

[1] DONDENA Centre for Research on Social Dynamics and Public Policy, Bocconi University

[2] Center for Health Emergencies, Bruno Kessler Foundation

[3] Department of Economics and Management, University of Pisa

[4] Department of Statistics, Computer Science, Applications, University of Florence

[5] Laboratory for Computational Epidemiology and Public Health, Department of Epidemiology and Biostatistics, Indiana University School of Public Health

[*] Co-first authors

[#] Joint senior authors

## Abstract

The collection of updated data on social contact patterns after the disruptions brought by the COVID-19 pandemic is of critical importance for future epidemiological evaluations and for the assessment of non-pharmaceutical interventions based on physical distancing.

Here, we run two waves of the same online survey on March 2022 and March 2023 in Italy, collecting information from a representative sample of the population on direct (verbal or physical interactions) and indirect (prolonged co-location in indoor spaces) contacts. We investigated the determinants of an individual's total number of social contacts using a generalized linear mixed model and assessed the potential impact of work-from-home and distance learning on the transmissibility of novel respiratory pathogens.

We found that in-person attendance to work or school was among the main drivers of social contacts: among adults (age 18 or more), in-person attendees reported a mean of 1.69 (95%CI: 1.56-1.84) times the contacts of adults staying at home; this ratio increased to 2.38 (95%CI: 1.98-2.87) among children and adolescents (less than 18 years old). We estimated that even drastic measures suspending all non-essential working activities, in the absence of any other preventive measures, would result in a marginal reduction of transmissibility. The combination of distance learning covering educational levels from the primary and upwards, with work-from-home for all jobs allowing for it, could reduce transmissibility by up to 23.7% (95%CI: 18.2-29.0%). Additionally suspending early childcare services would result in minimal further gains.

These results provide useful data for modelling the transmission of respiratory pathogens in Italy after the end of the COVID-19 emergency. They also provide insights on the potential epidemiological effectiveness of social distancing interventions centred on work and school attendance, supporting considerations on the balance between the expected benefits and their heavy societal costs.

## Introduction

Respiratory infectious diseases are spread in human populations through interactions associated to social contacts, occurring during the execution of everyday life activities. During severe health emergencies caused by respiratory infectious agents, such as pandemics, the limitation in the number of social contacts enacted through governmental mandates and recommendations (sometimes referred to as "social distancing" measures) can be an effective means to interrupt the chains of transmission and rapidly curb the disease spread [1,2]. However, such measures may come at extensive economic and societal costs. Among the many measures aimed at reducing the transmission of SARS-CoV-2, school closures have been widely debated[3] amidst concerns about their negative short- and long-term impact on young individuals' education and psychological well-being[3–5]. In addition to this, the suspension of most economic activities during the lockdown enacted by many countries in the spring of 2020 has resulted in massive economic damage, disruption of supply chains, and job losses[6], exacerbating poverty and inequality. In order to properly weigh the costs and benefits of social distancing measures, a crucial step is to quantify their impact on transmission. This difficult task is usually addressed through mathematical modelling [9-12,18-19], which largely relies on the availability of information on how many social contacts individuals experience across different socio-demographic stratifications, social settings and activities. Although such information was available, to some extent, for most countries at the onset of the COVID-19 pandemic[13-15], the individuals' behavioural responses and imposed restrictions[16–17] made it somewhat outdated. Several social contact studies were conducted in selected countries to assess changes in human mixing patterns throughout the course of the epidemics [12–16]. However, after COVID-19-related restrictions have progressively been lifted in most countries starting in 2022[17,18], only a few studies have evaluated post-pandemic contact patterns[10,19,20]. Here, we provide updated evidence on contact patterns in Italy obtained by internet-based surveys conducted in March 2022 and March 2023. We combine the collected contact pattern data with data on educational demographics and age-specific in-person workforce participation, to quantify the potential impact of targeted governmental restrictions on the transmissibility of a novel respiratory pathogen.

## Results

We conducted two waves of the same online social contact survey on a sample representative of the age, sex, and area of residence of the Italian population. The first wave covered the period from March 17 to March 29, 2022, and the second from March 17 to April 5, 2023. Participants were asked to provide their socio-demographic and health-related information and to report both direct and indirect contacts encountered on the day before the survey. A direct contact was defined as a person with whom physical interactions and/or in-person verbal exchanges of at least five words occurred. Participants were asked to provide details on each of their direct contacts, such as their age, sex, or the setting in which the interaction occurred. Participants were additionally asked to estimate the number of indirect contacts encountered on the previous day. An indirect contact was defined as a person with whom the respondent was co-located in a closed environment for at least 30 minutes (e.g., classmates at school, colleagues at work). For the main analysis in this study, we defined a contact as any direct or indirect contact (see Materials and Methods). The survey was administered to both adults (18 or more years old) and minors (less than 18 years old). Participants aged 14 to 17 years were requested to compile the questionnaire under the supervision of a legal guardian (usually a parent); for participants under 14, the compilation was performed by a legal guardian and the active engagement of the minor participant was strongly encouraged to ensure accurate responses to the questions.

The final sample included 3,743 participants producing 4,979 valid questionnaires, of which 2,474 were from the 2022 wave and 2,505 from the 2023 wave. A subset of 1,236 participants responded to both the 2022 and 2023 waves. Summary descriptions of the participants' characteristics and the corresponding mean numbers of contacts are reported in Table 1 and Table 2. Full details on the collected information are reported in the Supplementary Information.

The sample across the two waves included 127 children of preschool age (0-5 years), 64.4% of whom were enrolled in early childhood education services (ISCED level 0[21]), and 312 individuals aged between 6 and 17 years, all of which were enrolled in ISCED levels between 0 and 3. Among participants aged 18 or older (3304, representing 88.3% of respondents), 53.0% were employed either full-time or part-time, 7.7% were homemakers, 4.3% were students (full-time or part-time), 25.0% were retired, and 9.9% were inactive.

A total of 37,584 contacts were reported in the two study waves, of which 23,718 (63%) were direct and 13,866 (37%) were indirect (see Tables SI6, SI7, and SI8 for a detailed breakdown of direct and indirect contacts). Overall, 95.3% of the responses reported at least one contact on the previous day. Most of the contacts (90.8%) were reported in indoor settings, primarily at home (58.7% of total contacts). Contacts at school or work accounted for 16% of reported contacts, leisure social contacts for 13.1%, contacts on transportation means for 1.6%, and 10.6% of contacts were reported in other non-specified settings (see Table SI8).

The mean number of reported daily contacts per respondent across the two waves was 7.5 (95% bootstrapped Confidence Interval (CI) of the mean: 7.2-7.9) and was highly heterogeneous by age group, educational attainment and household size (Table 1). Significant differences were found with respect to employment status and attendance at work/school (Table 2). Among adult respondents who were employed, those who worked in person reported a higher number of social contacts, 9.4 (95%CI: 8.7-10.1), compared to those who worked remotely or did not work, 5.2 (95%CI: 4.6-6.0). Attendance to work was similar across the two waves (25.1% of the sample reported working remotely or not working in the 2022 wave and 26.5% in the 2023 wave). Similarly, students attending schools and universities in person reported a higher number of contacts, 16.2 (95%CI: 14.8-17.6), compared to those who did distance learning or did not attend school at all, 6.8 (95%CI: 6.5-7.2). During the 2022 wave, 73.6% of students attended school or university in person, compared to 86.1% among students from the 2023 wave.

### Determinants of social contacts

We modelled the total number of contacts as a function of covariates measured in the surveys using a generalised linear mixed effects model (GLMM) with negative binomial distribution and log-link function (see Methods and Sec. SI1.3). Two separate models were fitted to data from adults and minors, to account for the different available covariates (see Sec. SI2.1). In adults, the main factors contributing to a higher number of contacts were in-person attendance to work/school, living in larger households, younger age, and having completed the primary cycle (i.e., at least two doses) of the COVID-19 vaccine (Figure 1). Adults who attended school or work in person reported 1.69

(95%CI: 1.56-1.84) times the contacts of those who attended remotely or did not attend. Adults who lived in households of size 4 reported 1.50 (95%CI: 1.32-1.71) times the contacts of individuals living alone. Individuals of age 40 or more reported significantly fewer contacts than individuals in the age group 18-29, with a contact rate ratio ranging, on average, between 0.70 (95%CI: 0.60-0.81) in the age group 40-49 and 0.81 (95%CI: 0.69-0.95) in the age group 60-69. Respondents who received at least two COVID-19 vaccine doses reported 1.26 (95%CI: 1.11-1.42) times the contacts of unvaccinated or incompletely vaccinated respondents. Weaker but significant effects were attributed to educational attainment, income levels, and the presence of a cohabitant with underlying conditions. Small but significant differences were also found between first- and second-wave respondents, with the latter reporting 1.11 (95%CI: 1.06-1.17) times the contacts of the former. Consistently with results for the adult population, in-presence attendance to school was again the most important determinant of the number of contacts among minors, with in-person attendees reporting 2.38 (95%CI: 1.98-2.87) times the contacts of students not attending (see Figure SI4).

We ran a sensitivity analysis where we considered, for both minors and adults, the alternative outcome variable of direct contacts only. Results show that in-person school/work attendance was also in this case the most important determinant of the number of contacts (see Sec. SI2.3).

### Effect of in-person school and work attendance on viral transmissibility

Based on data collected in the two waves, we estimated the age-specific contact matrix for the Italian population (Figure 2A), and we disaggregated it by respondents who attended school or work in person (Figure 2B) and by those who did not (Figure 2C). Individuals who attended school or work in person are characterized by an overall higher number of interactions across all age groups, and by a higher level of assortative mixing in younger age groups (top panels in Figure 2). We used the contact matrices disaggregated by attendance in person to assess the impact of work-from-home and distance learning in reducing the transmission potential of a generic respiratory virus, in absence of other measures. We considered different scenarios combining various levels of in-person work and school attendance[21,22], and we quantified the relative change in reproduction numbers compared to a baseline scenario with full in-person attendance. Scenarios considered reflect the progressive physical closure of education levels, starting from tertiary education (ISCED 5-8) and progressively including lower levels, combined with three levels of in-person attendance at work: i) "pre-pandemic", corresponding to data before 2020; ii) "sustainable", reflecting the proportion of in-person workers observed in Italy after the reopening of economic activities following the COVID-19 lockdown on May 18, 2020; and iii) "minimum", where only essential workers are allowed to work in person, as during the 2020 national lockdown. The age-specific populations of students enrolled in the different education levels are reported in Figure 3A and those of in-person workers in the three scenarios are reported in Figure 3B.

We estimate that the maximum reduction in transmissibility compared to the baseline scenario using measures targeting only work attendance was 7.0% (95%CI: 4.8- 9.0%) (Figure 3C), obtained when only essential workers are allowed to attend in person. A "sustainable" scenario where only limited and sustainable work-from-home mandates were put in place on non-essential[22] resulted in a transmissibility reduction of only about 3%. Adding to this scenario the progressive suspension of in-person education levels would result in transmissibility reductions of: 5.0% (95%CI: 3.2-

6.8%) when tertiary education alone is suspended, (ISCED 5-8); 11.9% (95%CI: 7.8-16.4%) when including the upper secondary (ISCED 3); 16.2% (95%CI: 10.2-22.4%) with lower secondary (ISCED 2); 23.7% (95%CI: 18.2-29.0%) with primary education (ISCED 1); and 24.9% (95%CI:19.1-30.4%) with early childhood education (ISCED 0). The maximum transmissibility reduction, achievable by also limiting attendance in-person to only essential workers in addition to suspending all in-person schooling, was estimated to be 30.3% (95% CI: 22.6–35.9%).

## Discussion

In this study, we analysed data from a social contact survey conducted in Italy in two waves. The first wave took place in March 2022, immediately after the surge and decline of the Omicron variant[23,24]. The second wave was conducted in March 2023, several months after all restrictions had been lifted[25] and a month before the declaration of the end of the COVID-19 Public Health Emergency of International Concern. Respondents in the second wave reported a slight but statistically significant increase in the number of social contacts compared to the first wave, which may be attributed to the relaxation of the remaining restrictions and spontaneous behaviour change (see Figure SI17 for contact matrices differences between the two waves when accounting for in-person attendance).

Similarly to a previous study run in late 2022 in the UK, Belgium and the Netherlands, we found that the mean number of contacts in Italy in 2022/2023 has increased compared to those recorded in 2021[26–28] (Table SI3 and SI4). In line with other recent contact studies, we identified a positive association between the number of reported contacts and higher education levels, higher income, larger household size, lower age and COVID-19 vaccination[29–31].

Importantly, we found that the strongest determinant of the total number of contacts was in-person attendance at work or school. Other contact studies have found a similar relationship between workers and non-workers, but our peculiar focus on work-from-home and distance learning on the number of contacts allowed us to provide quantitative insights on potential governmental interventions acting on work/school in-person attendance. We quantified the reduction in transmission potential of a respiratory virus transmitted through direct contacts or shared closed spaces that could be allowed by combinations of school closures and working-from-home mandates (or the suspension of non-essential economic activities altogether). In agreement with previous findings, we found that suspending in-person education has a generally stronger impact on transmissibility (up to 20% when applied to all education levels in the absence of measures on work attendance)[32,33] than reducing in-person workforce (up to 7% when all non-essential productive sectors are suspended). A combination of both interventions at the maximum level (similar to what was implemented during the Italian lockdown in March-April 2020) is expected to contribute to a reduction of the transmissibility of up to 30% in the absence of other preventive measures such as mask use, isolation of diagnosed individuals, tracing and quarantining of contacts, ventilation of closed spaces, restrictions on other social contacts, and spontaneous protective behaviour. The suspension of non-essential economic activities always had a limited (<5%) additional effect on further reducing transmission when compared to sustainable work-from-home mandates.

Distance learning is a highly debated measure due to its implications on the quality of education and psychological well-being of children and young adults and on the increased burden on parents who need to rebalance childcare with their work responsibilities. In particular, closing lower educational levels poses stronger challenges to families, since young children engage less effectively in distance learning[34] and require a higher intensity of care. In this context, a relevant finding of this study is that maintaining in-person attendance for early childhood education (children aged 0-5 years) minimally affects the effectiveness of intervention. We acknowledge, however, that since questionnaires for young children were compiled by their legal guardians, they may be more prone to biases due to second-hand reporting of the number of contacts; therefore, we advise caution in the interpretation of this result.

In the context of social contact studies for epidemiological modelling, special attention needs to be placed on the definition of contacts. Traditional definitions, such as those proposed by Mossong et al.[7] or Coletti et al.[20], emphasise the importance of physical and conversational interactions. However, COVID-19 highlighted the importance of airborne transmission in shared indoor spaces[35–37]. To account for this, we included indirect contacts due to co-location in closed spaces within the definition of contact in the main analysis. One limitation of this choice is that details on the contact (e.g., their age) would be difficult to recall for all co-located contacts (for example, customers attending the same restaurant or bar) given the lack of personal interaction and their potentially large number. Therefore, we chose not to collect information details on indirect contacts to avoid long survey completion times resulting in high dropout rates or low-quality responses. Information on the age of indirect contacts was inferred based on the details provided for direct contacts. Nonetheless, a sensitivity analysis where we considered only direct contacts provided consistent results with with those of the main analysis (reported in the Supplementary Information).

The impact of work-from-home and distance learning measures estimated here considered the spread of a generic pathogen through close contacts and airborne transmission in a fully susceptible population, neglecting, for example, possible age-specific heterogeneities in susceptibility and infectiousness. As such, results may need to be recalibrated to the features of the actual pathogen for which such measures are taken in consideration. However, data collected and made available through this study make such a reassessment relatively straightforward once the characteristics of the pathogen are known.

Although we believe that qualitative results from this study may hold for countries with similar socio-demographic and economic structures, we caution against their direct extrapolation to other geographical settings, given heterogeneities in educational systems, workplace structures and social interactions at work, household composition, age-specific population size and contact patterns.

This research provides data across multiple sociodemographic strata on social contact patterns in Italy after the COVID-19 emergency, and analytics identifying the effect of socio-economic and demographic determinants on the number of experienced social contacts. These results were combined with data on school enrolment and in-person work attendance to provide estimates of the potential impact of public health interventions involving the educational and productive sector. Estimates of this kind can support considerations on the balance between the expected epidemiological benefits and their societal costs.

## Materials and Methods

### Data and data cleaning

This work is based on a new sample of data collected online in two waves at the end of March 2022 and 2023, from a nationally representative panel of the Italian population in terms of age, sex, and area of residence. Except for minor updates (e.g. for the number of vaccine doses recommended to the population in 2023), the second wave was identical to the first and administered to a random group of respondents who already participated to the first wave (1,246 participants), and to a group of first-time respondents of approximately equal size.

The survey consists of two main sections: i) socio-demographics, health-related information and behavioural information on the respondent, and ii) a contact diary in which respondents were asked to recall their direct and indirect contacts on the day prior the survey administration. After data collection, responses underwent an accurate data-cleaning process including: i) removal of respondents with inconsistencies in responses, and ii) removal of respondents with incomplete information. A complete description of the data-cleaning procedure is provided in Sec. SI1.1.

### Indirect co-location events augmentation procedure

For each direct contact, additional information was collected about the characteristics of the interaction and information on contacts themselves. However, indirect contacts were only reported as an aggregate number. To run the statistical model and construct age-specific contact matrices it was thus necessary to augment indirect contact data with the relevant missing information.

The data augmentation procedure used in the baseline analysis can be summarized as follows. For each respondent:

i) we defined the number of indirect contacts to be augmented (N), by subtracting from the reported count for indirect contacts the number of indoor direct contacts that they reported. This assumption conservatively accounts for contacts potentially reported both as direct indoor and as indirect.

ii) we reconstructed the age and setting of indirect contacts by sampling N times (with replacement) this information from the set of direct contacts reported by the same participant, excluding cohabitants and outdoor contacts.

iii) if the respondent did not report direct contacts, indirect contacts data augmentation was performed by sampling the information from direct contacts of other participants of the same age as the respondent.

We performed a sensitivity analysis on the data augmentation procedure by limiting the sampling for the assignment of indoor contacts only to the setting in which the respondent reported most of their direct contacts, rather than sampling from the overall direct contacts. See Section SI1.2 for more details.

## Determinants of social contacts: the statistical framework

To investigate the determinants of the total number of contacts, we fitted a generalised linear mixed model[38]. We initially screened 35 covariates relative to the respondent and cohabitant's characteristics for significant associations with the total number of contacts, fitting a set of negative binomial regression models having as independent variables each individual covariate, an intercept, and terms accounting for the wave (2022 or 2023) and for whether the respondent participated to both waves or not (see Table SI5 for a full list of the screened covariates). Covariates that were not significant in the multivariate model were subsequently filtered out, and the final model included 15 covariates: sex, age group aggregated by 10-year intervals, occupation, household income, household size, in-presence attendance at work/school, contact happened on a Sunday, completion of SARS-CoV-2 primary vaccination cycle, recent (in the last 4 months) SARS-CoV-2 infection, presence of chronic comorbidities in the respondent, presence of chronic comorbidities in cohabitants, senior (65 years or older) cohabitants, having children, and participation to the first survey wave. Finally, a random intercept accounted for within-individual correlation in longitudinal responses across the two waves. Full details on the generalised linear mixed model selection are reported in SI1.3.

## Age-specific contact matrices

We constructed age-specific matrices of average total contacts for the Italian population stratified by 15 age groups (0-4, 5-9, 10–14, 15-19, 20-24, 25-29, 30-34, 35-39, 40-44, 45-49, 50-54, 55-59, 60-64, 65-69, and 70+ years old). We then computed the population normalised reciprocal contact matrix, following established procedures from the literature[7,14,19]. The matrices for the average total contacts obtained by 1000-fold bootstrapping of responses from both waves are reported in Figure 2 (see also Figures SI9 and SI11 for robustness). In the SI, we report contact matrices disaggregated by setting (Figures SI5-7) and matrices including only direct social contacts (Figure SI9).

## Effect of school and work attendance on the transmissibility of a generic respiratory virus

We evaluated the potential effect of different scenarios of school and work attendance on the transmissibility of a novel respiratory virus (i.e., spreading in a fully susceptible population in the absence of interventions) under the assumption of homogeneous susceptibility and infectiousness across age groups. To do so, we combined data from Eurostat on age-specific populations attending different education levels[21] (from early childhood to tertiary education) and on age-specific in-person workforce in Italy from a previously published study[22].

We denote by $R_s$ the reproduction number associated with scenario $s$ and compute the relative reduction $\alpha_s$ with respect to the baseline scenario with full school and work attendance:

$$\alpha_s = 1 - \frac{R_s}{R_0} = 1 - \frac{\rho(NGM_s)}{\rho(NGM_0)}$$

where $\rho(NGM_s)$ and $\rho(NGM_0)$ represent the dominant eigenvalues of the next-generation matrices associated with scenario $s$ and with the baseline scenario, respectively [22,39,40]. Since infection-related parameters are the same across scenarios, $\alpha_s$ can be simplified to:

$$\alpha_s = 1 - \frac{\rho(M_s)}{\rho(M_0)}$$

where $M_s$ is a block-matrix

$$M_s = \begin{pmatrix} A_{i,j}^s & A_{i,j}^s \\ B_{i,j}^s & B_{i,j}^s \end{pmatrix}$$

The blocks are defined as

$$A_{i,j}^s = C_{i,j}^P \frac{N_i^P(s)}{N_j^P(s) + N_j^{NP}(s)} \qquad \text{and} \qquad B_{i,j}^s = C_{i,j}^{NP} \frac{N_i^{NP}(s)}{N_j^P(s) + N_j^{NP}(s)}$$

with

- $C_{i,j}^P$ representing the overall contact matrix estimated for the population attending work or school in-person;
- $C_{i,j}^{NP}$ representing the overall contact matrix estimated for the population not attending work or school in-person;
- $N_i^P(s)$ representing the number of individuals of age $i$ attending schools or work in-person in scenario $s$;
- $N_i^{NP}(s)$ representing the number of individuals of age $i$ not attending schools or work in-person in scenario $s$.

## Ethics statement

This study was conducted in accordance with the ethical standards set by Bocconi University and has received approval from the Bocconi University Ethical Board (Approval Number: FA000383 - 17 January 2022).

Participants were informed about the nature and purpose of the research, including the voluntary nature of their participation and their right to withdraw at any time without any negative consequences. Informed consent was obtained from all participants prior to their involvement in the study. All analyses were carried out on anonymized data.

## Acknowledgment & Funding

LL, CC, FT, VO, EDA, AM acknowledge funding from the ERC Consolidator Grant IMMUNE (no. 101003183). Researchers from the Bocconi Covid Crisis Lab acknowledge funding from the Romeo and Enrica Invernizzi Foundation. VM, GG, MM, PPo, and SM acknowledge funding from the Fondazione Valorizzazione Ricerca Trentina (VRT), project COVIDVAX. This research was supported by EU funding within the NextGeneration EU-MUR PNRR Extended Partnership initiative on Emerging Infectious Diseases (Project No. PE00000007 INF- ACT).

# Figures and Tables

*Table 1: Survey respondents' characteristics and average contacts by wave.* The table reports the number of respondents (and the percentage with respect to the number of respondents in each wave) and respondents' average number of social contacts (and 95% credible interval) stratified by different respondent characteristics: age, sex, household income, and household size. Average and credible intervals (CrI) are computed through 10,000 bootstrap iteration sampling with replacement from the set of respondents within each demographic group.

| Variables | March 2022 wave | | March 2023 wave | | Both waves | |
|---|---|---|---|---|---|---|
| | Sample size | Avg. Contacts (95% CI) | Sample size | Avg. Contacts (95% CI) | Sample size | Avg. Contacts (95% CI) |
| **Total** | 2474 | 7.3 (6.9 - 7.8) | 2505 | 7.8 (7.4 - 8.2) | 4979 | 7.5 (7.2 - 7.9) |
| **Age group** | | | | | | |
| 0-9 | 150 (6.1%) | 11.1 (9.5 - 12.7) | 112 (4.5%) | 12.6 (10.5 - 14.8) | 262 (5.3%) | 11.7 (10.5 - 13) |
| 10-19 | 165 (6.7%) | 12.7 (10.9 - 14.6) | 103 (4.1%) | 15.2 (12.5 - 18) | 268 (5.4%) | 13.6 (12.1 - 15.2) |
| 20-29 | 202 (8.2%) | 10.4 (8.5 - 12.6) | 185 (7.4%) | 10.8 (8.8 - 13.1) | 387 (7.8%) | 10.6 (9.2 - 12.2) |
| 30-39 | 376 (15.2%) | 8 (6.9 - 9.2) | 357 (14.3%) | 8.8 (7.6 - 10.1) | 733 (14.7%) | 8.4 (7.6 - 9.3) |
| 40-49 | 381 (15.4%) | 6 (5.1 - 7.1) | 339 (13.5%) | 6.7 (5.8 - 7.7) | 720 (14.5%) | 6.4 (5.7 - 7.1) |
| 50-59 | 432 (17.5%) | 6.3 (5.4 - 7.2) | 506 (20.2%) | 7.4 (6.4 - 8.3) | 938 (18.8%) | 6.9 (6.2 - 7.5) |
| 60-69 | 320 (12.9%) | 6.5 (5.4 - 7.7) | 358 (14.3%) | 6.3 (5.3 - 7.4) | 678 (13.6%) | 6.4 (5.6 - 7.2) |
| 70+ | 448 (18.1%) | 4.8 (4.2 - 5.5) | 545 (21.8%) | 5.7 (5 - 6.5) | 993 (19.9%) | 5.3 (4.8 - 5.8) |
| **Sex** | | | | | | |
| Male | 1248 (50.4%) | 7.5 (6.9 - 8.1) | 1272 (50.8%) | 8.2 (7.6 - 8.9) | 2520 (50.6%) | 7.9 (7.4 - 8.3) |
| Female | 1226 (49.6%) | 7.1 (6.6 - 7.8) | 1233 (49.2%) | 7.3 (6.8 - 7.9) | 2459 (49.4%) | 7.2 (6.8 - 7.6) |
| **Net household income (euros per month)** | | | | | | |
| <1,500 | 583 (23.6%) | 5.8 (5.1 - 6.7) | 568 (22.7%) | 6.3 (5.5 - 7.2) | 1151 (23.1%) | 6.1 (5.5 - 6.7) |
| 1,500-2,999 | 967 (39.1%) | 7.5 (6.8 - 8.2) | 1002 (40%) | 7.6 (7 - 8.2) | 1969 (39.5%) | 7.5 (7.1 - 8) |
| >3,000 | 509 (20.6%) | 8.9 (8 - 10) | 551 (22%) | 10.2 (9.1 - 11.5) | 1060 (21.3%) | 9.6 (8.8 - 10.4) |
| Undisclosed | 415 (16.8%) | 7.1 (6.2 - 8.2) | 384 (15.3%) | 6.8 (5.8 - 7.8) | 799 (16%) | 7 (6.3 - 7.7) |
| **Household size** | | | | | | |
| 1 | 357 (14.4%) | 5.2 (4.5 - 6) | 355 (14.2%) | 5.9 (4.9 - 6.9) | 712 (14.3%) | 5.5 (4.9 - 6.2) |
| 2 | 745 (30.1%) | 5.7 (5 - 6.5) | 844 (33.7%) | 6.5 (5.7 - 7.2) | 1589 (31.9%) | 6.1 (5.6 - 6.6) |
| 3 | 732 (29.6%) | 8.2 (7.3 - 9.1) | 677 (27%) | 8.5 (7.6 - 9.3) | 1409 (28.3%) | 8.3 (7.7 - 8.9) |
| 4 | 502 (20.3%) | 9.1 (8.2 - 10.1) | 492 (19.6%) | 9.6 (8.7 - 10.6) | 994 (20%) | 9.3 (8.7 - 10.1) |
| 5+ | 138 (5.6%) | 10.5 (8.7 - 12.6) | 137 (5.5%) | 11.1 (9.3 - 13.1) | 275 (5.5%) | 10.8 (9.4 - 12.2) |
| **COVID-19 vaccination status** | | | | | | |
| None | 218 (8.8%) | 6.8 (5.7 - 8.1) | 212 (8.5%) | 6.4 (5.3 - 7.6) | 430 (8.6%) | 6.6 (5.8 - 7.5) |
| 1 dose | 23 (0.9%) | 7.2 (4.6 - 10.4) | 28 (1.1%) | 7.9 (5.1 - 11) | 51 (1%) | 7.6 (5.6 - 9.7) |
| 2 doses | 290 (11.7%) | 8.7 (7.4 - 10.2) | 310 (12.4%) | 8.6 (7.4 - 10) | 600 (12.1%) | 8.7 (7.7 - 9.7) |
| 3+ doses | 1839 (74.3%) | 7 (6.6 - 7.6) | 1880 (75%) | 7.6 (7.1 - 8.1) | 3719 (74.7%) | 7.3 (7 - 7.7) |
| Exempt | 104 (4.2%) | 9.3 (7.6 - 11.2) | 75 (3%) | 13 (9.8 - 16.7) | 179 (3.6%) | 10.9 (9.2 - 12.8) |

**Table 2:** *Survey respondents' sample and average contacts by wave.* The table reports the number of respondents included in the survey (and their population percentage with respect to the number of respondents per wave) and respondents' average number of social contacts (and 95% credible interval) stratified by additional respondent characteristics: employment status, working mode (e.g. remotely or in-presence), teaching method (e.g. distance or in-presence learning), COVID-19 vaccinal status, and COVID-19 verified infections. Average and Credible Intervals (CrI) are computed through 10,000 bootstrap iteration sampling with replacement from the set of respondents within each variable group.

| Variables | March 2022 wave | | March 2023 wave | | Both waves | |
|---|---|---|---|---|---|---|
| | Sample size | Avg. Contacts (95% CI) | Sample size | Avg. Contacts (95% CI) | Sample size | Avg. Contacts (95% CI) |
| **<u>Educational attainment</u>** | | | | | | |
| Lower secondary or below | 324 (13.1%) | 6.5 (5.4 - 7.6) | 329 (13.1%) | 6.5 (5.4 - 7.6) | 653 (13.1%) | 6.5 (5.7 - 7.2) |
| Upper secondary | 1380 (55.8%) | 6.8 (6.3 - 7.4) | 1377 (55%) | 7.3 (6.8 - 7.8) | 2757 (55.4%) | 7.1 (6.7 - 7.4) |
| Tertiary or above | 770 (31.1%) | 8.5 (7.8 - 9.4) | 799 (31.9%) | 9.2 (8.3 - 10.1) | 1569 (31.5%) | 8.9 (8.3 - 9.5) |
| **<u>Employment status</u>** | | | | | | |
| Employed (full or part-time) | 1103 (44.6%) | 7.6 (7 - 8.3) | 1199 (47.9%) | 8.5 (7.9 - 9.3) | 2302 (46.2%) | 8.1 (7.6 - 8.6) |
| Looking after home/family | 179 (7.2%) | 4 (3.4 - 4.6) | 204 (8.1%) | 5.8 (4.9 - 7) | 383 (7.7%) | 5 (4.4 - 5.6) |
| Student (full- or part-time) | 378 (15.3%) | 12.6 (11.3 - 13.9) | 253 (10.1%) | 14 (12.3 - 15.8) | 631 (12.7%) | 13.1 (12.1 - 14.2) |
| Retired | 272 (11%) | 5.6 (4.4 - 7.1) | 216 (8.6%) | 4.9 (4.1 - 5.8) | 488 (9.8%) | 5.3 (4.5 - 6.2) |
| Inactive | 542 (21.9%) | 5 (4.3 - 5.8) | 633 (25.3%) | 5.4 (4.8 - 6.1) | 1175 (23.6%) | 5.2 (4.8 - 5.7) |
| **<u>Working mode (only employed respondents during weekdays)</u>** | | | | | | |
| Remote or did not attend | 195 (25.1%) | 4.9 (4.1 - 5.7) | 218 (26.5%) | 5.6 (4.6 - 6.8) | 413 (25.8%) | 5.2 (4.6 - 6) |
| In-person | 582 (74.9%) | 8.5 (7.7 - 9.4) | 604 (73.5%) | 10.2 (9.2 - 11.3) | 1186 (74.2%) | 9.4 (8.7 - 10.1) |
| **<u>School attendance (only student respondents during weekdays)</u>** | | | | | | |
| Remote or did not attend | 72 (26.4%) | 8.4 (5.9 - 11.3) | 29 (13.9%) | 7.8 (5.2 - 11) | 101 (21%) | 8.2 (6.2 - 10.5) |
| In-person | 201 (73.6%) | 15.7 (14.1 - 17.4) | 179 (86.1%) | 16.8 (14.6 - 19.1) | 380 (79%) | 16.2 (14.8 - 17.6) |

**Figure 1:** *Rate ratios for the number of contacts in the adult population associated with respondent's characteristics.* Rate ratios higher (lower) than one indicate an increased (decreased) average number of contacts for the given covariate relative to the reference. Points: mean rate ratio; lines: 95%CI.

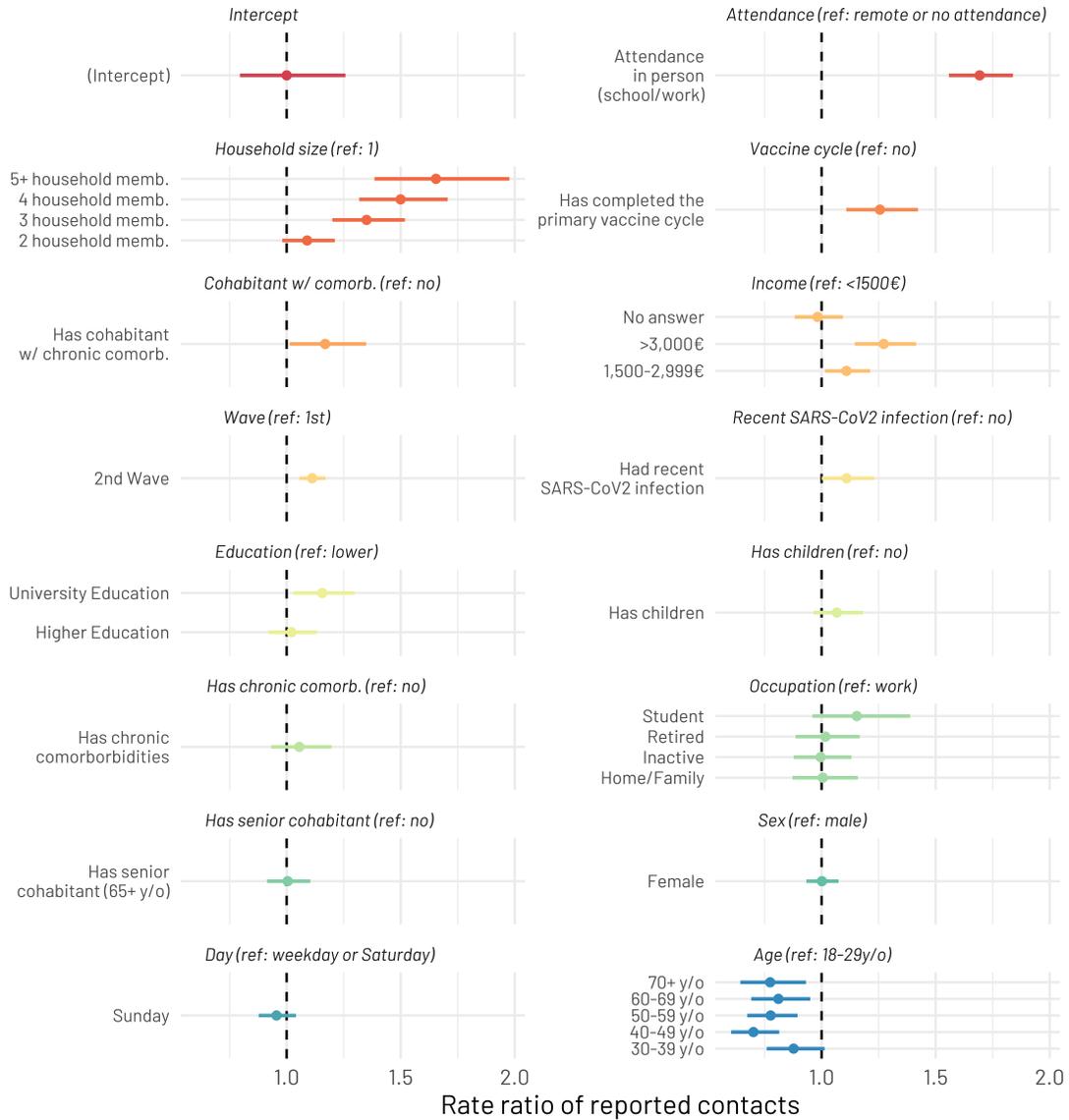

***Figure 2:*** *Age-specific contact patterns based on survey data from both waves.* Matrices were constructed by averaging 1,000 bootstrap samples with replacement of survey responses. Each cell represents the mean number of total contacts per individual, stratified by age group pairs. Top panels indicate the mean number of contacts (blue line) and the mean age assortativity[14,41] (red line) for each age group. A) Contact patterns obtained from the full sample, representative of the Italian population. B) Contact patterns for the subset of individuals attending school or work in person. C) Contact patterns for the subset of individuals not attending schools or working in person.

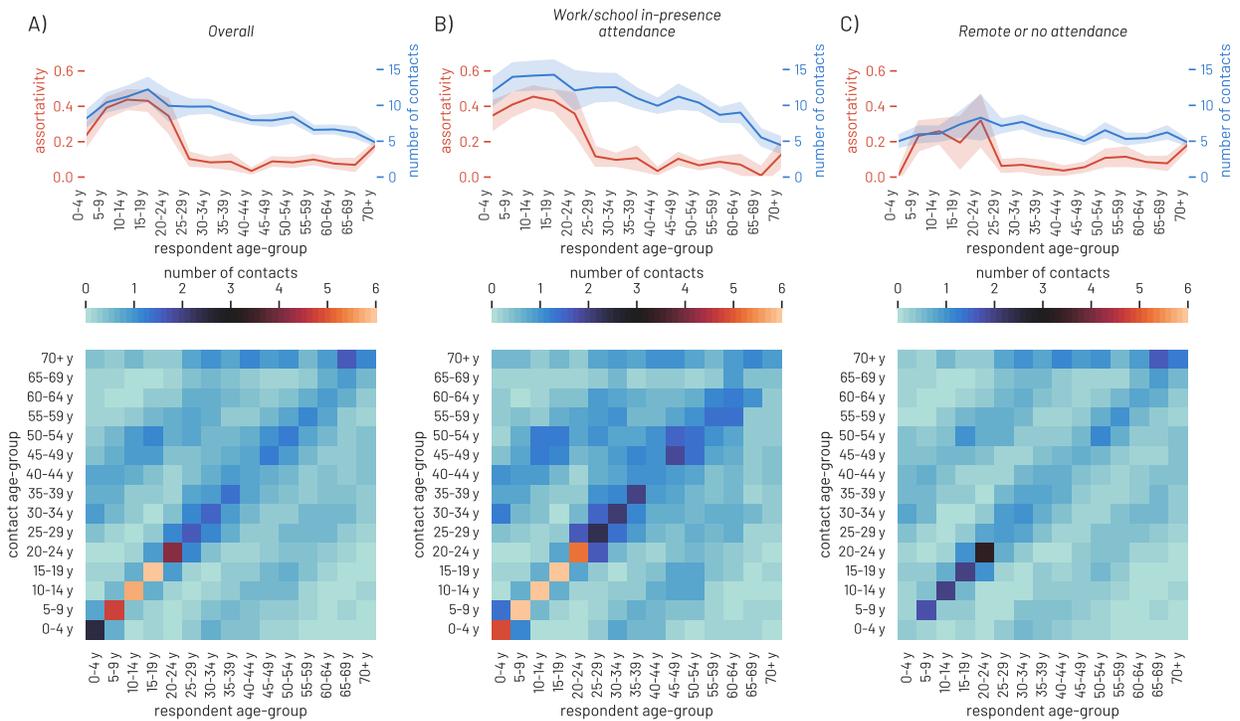

**Figure 3:** *Transmissibility reduction under different scenarios of work/school attendance.* A) Age-specific population enrolled to different educational levels in Italy in 2020, according to the ISCED classification 2011[21]. B) Age-specific population working in person in Italy under the three considered scenarios [20]: baseline (corresponding to pre-pandemic estimates); sustainable (corresponding to the population after the reopening of all economic activities following the end of the COVID-19 lockdown in May 2020); minimum (corresponding to in-person workers estimated during the COVID-19 lockdown]). C) Relative reduction in transmissibility associated with intervention scenarios combining different levels of school closures (including distance-learning mandates) and in-person work, compared to a baseline scenario where the in-person workers are set at pre-pandemic levels[22] and students across all education levels attend in person.

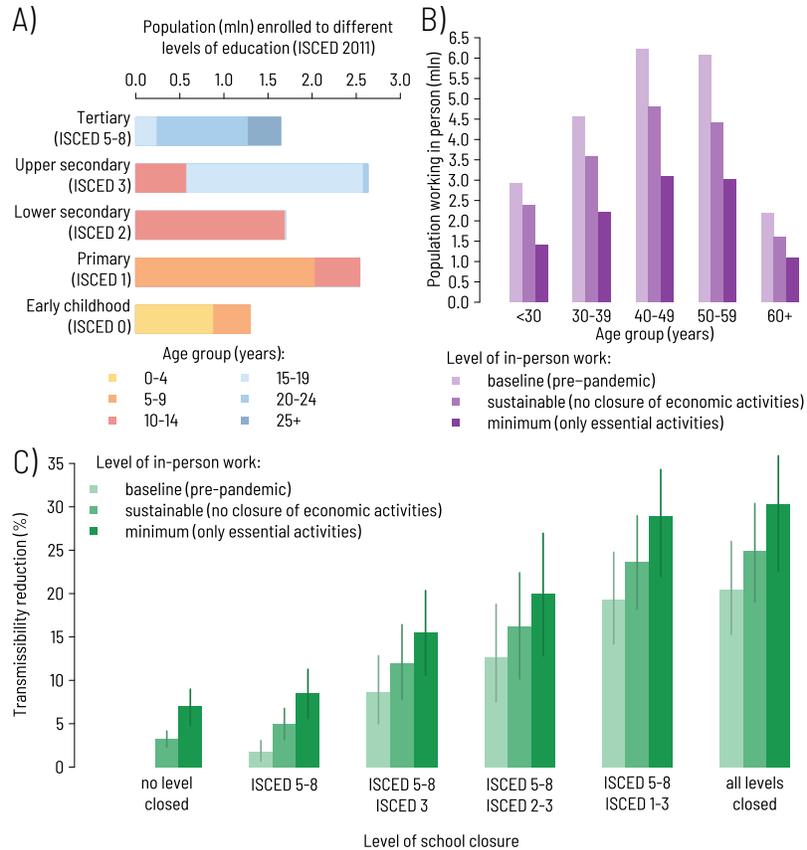

# Supplementary Information
## *"Social contact patterns in Italy after the COVID-19 emergency: implications for social distancing measures centred on in-person school and work attendance"*

## Table of Contents

# 1. Materials and Methods

## 1.1 Data

The analysis presented in this manuscript is based on a novel social contact survey data collection. The data collection consisted of two waves: the first run in the last two weeks of March 2022, the second run in the last two weeks of March 2023 and the first of April 2023. The collection consisted of about 3.500 responses for each of the two waves. The survey was designed to include both a panel and a longitudinal set of responses: about half of the responses collected in the second wave were obtained from respondents who had already participated in the first wave. In this section, we provide details about the survey design and data cleaning procedure that was adopted in the study.

### Survey methodology

We engaged the market research firm YouGov to carry out a survey on a representative sample of the Italian population in terms of age, sex, and NUTS level-1 area of residence. Members of YouGov's existing online panel (adults aged 18 and older) were invited to participate through email invitations. Adults with underage children were randomly asked to administer the survey to one of their children or involve them in the completion process. Children's involvement differed based on their age: those aged 14 to 17 were asked to directly respond to the survey, with their parent's support; legal guardians of children aged less than 14 years were asked to compile the survey themselves and request their children's feedback, e.g., on "if" and "with whom" their children had social contacts.

Participants were asked to provide socio-demographic information (such as age, gender, household income, educational attainment, and occupation) as well as health-related information (such as their COVID-19 vaccination status, whether they had had a positive SARS-CoV-2 test and when, or if they suffered from chronic or acute diseases at the time of data collection). Additionally, the survey collected information about the respondents' cohabitants (such as age, chronic diseases, or acute diseases of any cohabitants).

Differently from questionnaires administered for past social contact surveys, to facilitate recall of contacts and to reduce the impact of respondents' fatigue on the number of reported contacts, participants were asked first to report only a nickname for each direct contact, being guided to recall contacts for each setting: home (cohabitants or other contacts at home), work, school, restaurants, leisure places, transportation means, and other places). Only after the collection of all nicknames, respondents were asked to report information about each of their contacts and how the interaction occurred (e.g. the age and sex of the contact, whether it occurred in an open or closed setting, whether there was any physical interaction, the type of relation with the contact, the contact's COVID-19 vaccination status, their perceived income, who was wearing a mask during the interaction etc.). At the end of the diary, respondents were also asked to report the overall number of persons with which they shared an indoor space for a period of more than 30 minutes, including in the number people already reported as direct contacts (e.g. schoolmates or co-workers or people in public transport means with which they might not have sustained a direct conversation).

The survey was tested on a pilot sample of respondents to control for completion time and flow. Results from the analysis of the pilot responses reported a median completion time of about 15 minutes. Post-collection analysis confirmed similar completion times for both waves.

### Data cleaning

Before proceeding with the analysis reported in the main manuscript, we performed four main data-cleaning steps. A flowchart of the steps we performed is available in Figure SI11.

In the first data-cleaning step, we tackled issues that could affect the reliability of contacts reported by some respondents, e.g. by labelling unreliable respondents those students reporting to have attained to school during school closure periods (e.g. Sunday). Additionally, we identified four main categories of potential contact-specific problems that also resulted in the exclusion of the respondent as deemed "unreliable":

- **Duplicated contacts**: The online survey platform applied checks to prevent the respondents from writing the same nickname twice in order to avoid double-reporting of the same contact across different settings. Nevertheless, some respondents bypassed the quality check. Contacts for which respondents listed the same nickname-ages bundle across locations were considered duplicate contacts. The nickname check was bypassed by respondents in two ways:
  - The duplicates were listed in the "cohabitants" section (for which no check was implemented) and in at least another location section.
  - The duplicates are listed in different locations, with slight modifications of the nickname (e.g., "Alice a friend" and "Alice, a friend", or "Mother" and "mother").
- **Multiple contacts per entry**: some respondents reported more than one contact per line, using plurals, collective nouns, or providing the number of contacts (e.g., "10 friends", "parents", "clients").
- **Settings and contexts**: some respondent reported the names of the context or setting where they encountered contacts (e.g., "the supermarket", "Football"). This issue particularly affected contacts in crowded situations or where they could not easily be listed by the respondent.
- **Other issues**: other problematic entries which could not be otherwise classified. For instance, some respondents listed themselves among the contacts or reported "no one". Others wrote sentences by using the different available lines, e.g., to express they did not want to disclose them (e.g., "no" – "don't" – "declare").

Note that, answers from a single respondent were possibly affected by more than one issue at the same time.

At the end of this process, we excluded 2339 responses for which at least one entry was considered problematic. In the case of recontacts, only the problematic response among the two waves completed was deleted. We stress that this approach may have increased the relative number of individuals reporting 0 contacts, as the response quality of individuals reporting 0 contacts could not be tested following this schema, thus leaving their overall number unaltered.

In the second step, we cleaned the dataset from responses declaring the adoption of any kind of isolation measure (whether mandatory or voluntary) on the previous day, or for which the information was missing.

In the third step, to run the models, we dropped all non-complete cases based on all the variables we decided to include in the models themselves, i.e. cases for which at least one of the 35 required variables was missing. We also excluded 49 respondents among those who reported more than 100 contacts and children respondents reporting 0 contacts.

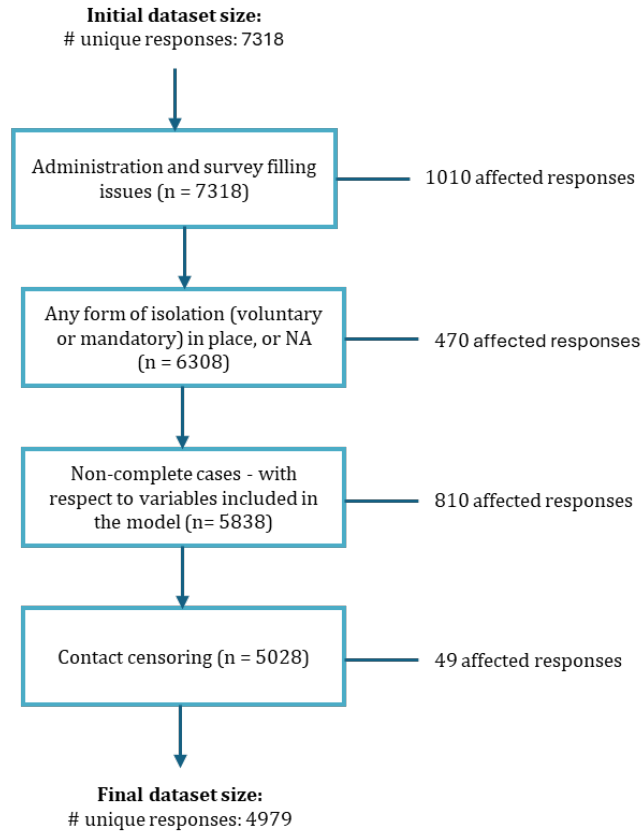

**Figure SI1**: *Dataset cleaning procedure flowchart*. This chart quantifies the impact of each cleaning step on the dataset size. On the right, at each step, we report the number of remaining responses that are affected by the addition of the cleaning requirement. The entire cleaning procedure reduces the number of responses from 7318 to 4979.

## Data representativeness

In this section, we present official 2022 data from the European Union Statistical Office (EUROSTAT) alongside to the characteristics of survey respondents who successfully completed the entire survey and passed all the quality checks described in the previous sections. Figure SI2 shows the fraction of respondents by age group (light blue bars), and the corresponding National Statistical Office estimates (red bars). The reported figure shows a good population representativeness by age, with the most significant differences between the expected population and the respondent population found in the underage population. Figure SI3 reports pair-wise combinations of age, area of residence, and gender for the respondent and expected distributions. Full-coloured bars show the fraction of each population group within the survey respondents panel while empty-red bars report the corresponding European Union Statistical Office estimates for the Italian population. Age distribution stratified by area of residence is highly representative of the expected population distribution by age (as shown in Fig. SI2). Similarly, a gender-based stratification shows an underrepresentation of the female population group in the highest age group. Education shows low representativity for the "lower secondary and below" educational attainment group. This result is in line with the expectations for online surveys, which systematically shows a lower level of participation from this population group.

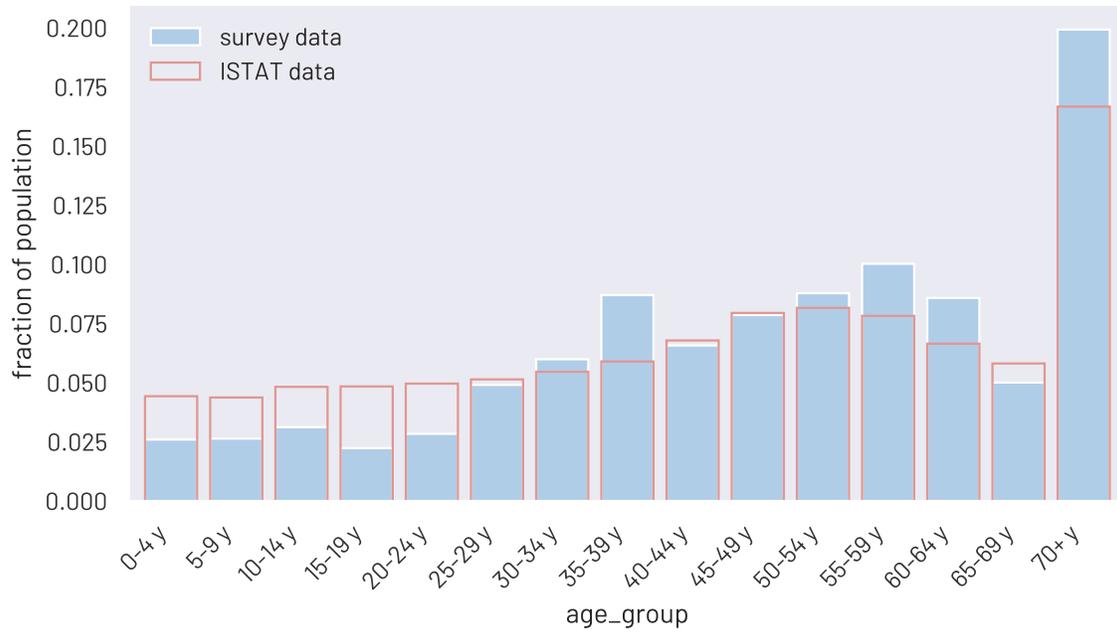

**Figure SI2:** *Survey respondents by age against the Italian population.* Light blue bars report the distribution of respondents by age group. Darker blue bars show the distribution by age of the respondents' reported contacts. The overlayed red bars report the fraction of the population by age group reported for 2022 by the Statistical Office of the European Union.

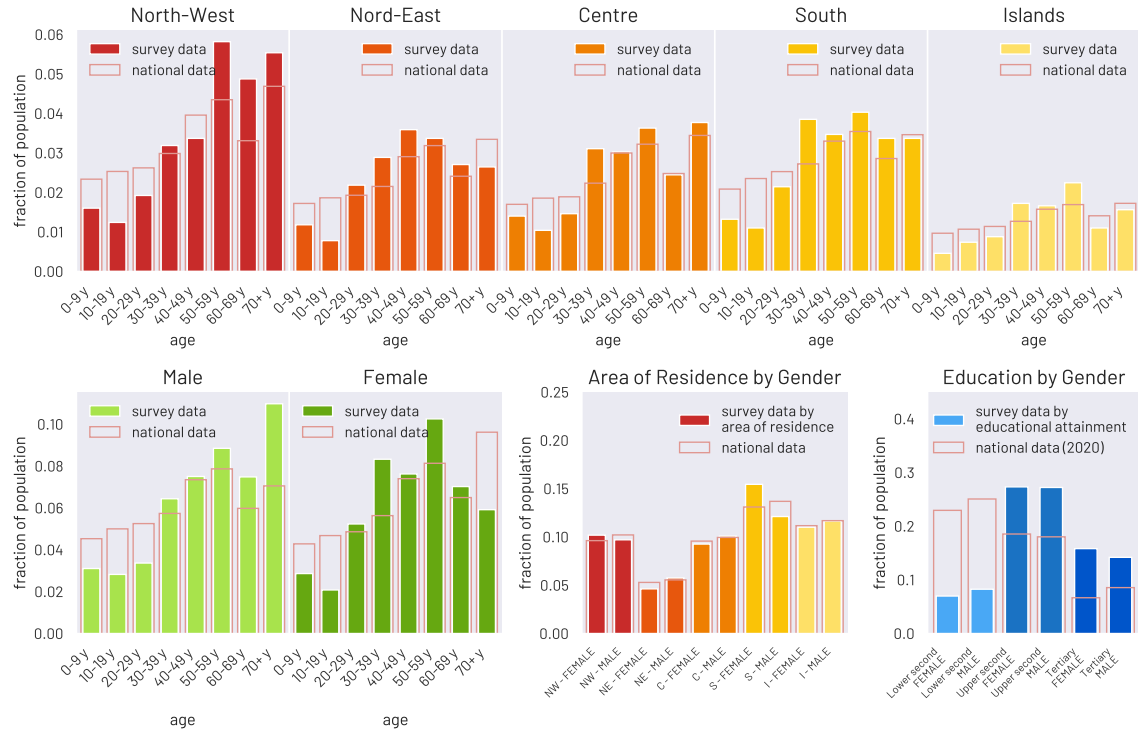

**Figure SI3:** *Survey respondents by age and area of residence, age and gender, area of residence and gender, and educational attainment and gender against the Italian population.* Full-coloured bars show the distribution of the survey respondents; overlayed red bars report the corresponding values reported for 2022 by the Statistical Office of the European Union. Reference national values for educational attainment by gender are sourced for 2020 from the Italian National Statistical Office

## 1.2 Indirect contacts augmentation

As discussed in the main manuscript, the analyses performed in this work consider both direct contact information, as defined by traditional social contact surveys, and information about indirect contacts, defined as events of a minimum duration of 30 minutes, occurring in a closed environment (such as a house, restaurant, transportation means, etc.). Indirect contacts were collectively reported in the form of a single numerical estimate without any additional information, e.g., about the setting in which they occurred or the contacts' characteristics. Thus, to compute social contact matrices that also include direct contacts we augmented the available information by: i) determining how many of the indirect contacts were already reported as direct social contacts , and ii) assigning to indirect contacts information on where and with whom the interaction happened.

### Direct and indirect contacts overlap

Respondents were requested to include in the estimate of indirect contacts also direct contacts that satisfied the indoor contacts' definition. A lower and an upper bound for the number of non-overlapping indirect contacts can be obtained by considering the two extreme cases:

1. all indoor direct contacts shared an indoor space for more than 30 minutes with the respondent.
2. none of the indoor direct contacts shared a space with the respondent for more than 30 minutes.

For example, if the reported direct contacts were N=6, of which 2 occurred indoors with cohabitants, 2 indoors at work, and 1 indoor at the restaurant or other leisure places (resulting in X=5 direct indoor contacts), and the indirect (indoor) contacts reported were Y=21, the two scenario would translate into i) having a total number of social contacts equals to 21 (*scenario 1*, with X direct indoor contacts and Y-X additional indirect contacts), or ii) having a total number of social contacts equals to 26 (*scenario 2*, with X direct indoor contacts and all the Y reported indirect contacts).

### Where and with whom?

To effectively use the indirect contacts reported by a respondent, additional contact information had to be inferred. Two alternative approaches were considered in this work to assign setting and contact information to indirect contacts.

A. The first assumes that indirect contacts had similar characteristics as direct contacts and were encountered in the same types of settings. In this framework, if a respondent reported X direct indoor contacts and Y indirect contacts, the information on the Y indirect contacts would be augmented by sampling directly from the reported X direct indoor contacts, excluding contacts with cohabitants and those encountered outdoors.

Using the example from the paragraph above, contact information would have been sampled 16 times in contact *scenario 1*. In this case, augmentation would be done by randomly sampling contact information from all indoor direct contacts (resulting, e.g., in 14 indirect contacts sampled from work-related direct contacts, and 7 contacts sampled from indoor leisure-related contacts).

B. The second approach assumes that all indirect contacts occurred in the setting where the majority of direct contacts were reported. Using the example above under *scenario 1*, this assumption translates into sampling 16 indirect contacts using only contact information from the indoor work setting.

**Table SI1:** *Description of the different scenarios for indirect contact inclusion.*

| | **Including indirect contacts** | |
| | Scenario 1 *Increment social contacts with Y-X additional indirect contacts* | Scenario 2 *Increment social contacts with all Y indirect contacts* |
|---|---|---|
| **Scenario A** *Sampling proportionally to indoor settings* | **Scenario 1A** *(Main manuscript analysis)* | **Scenario 2A** |
| **Scenario B** *Sampling form dominant indoor setting* | **Scenario 1B** | **Scenario 2B** |

(Row group label: **Enriching indirect contacts**)

In the main analysis, the total number of contacts considered direct and completely-overlapping indirect contacts where contact information augmentation was performed proportionally to the setting (scenario 1A).

Sensitivity analyses were run for alternative assumptions (scenario 1B and scenario 2A) to assess their impact on the statistical model, the contact matrices, and the transmissibility reduction. Results are reported in section 2 of this Appendix.

### 1.3 Statistical Framework

In this section, we report detailed information about the model selection procedure and the covariate screening procedure.

We model individual number of contacts using a negative binomial distribution. Negative binomial distribution is the standard choice to model count data with potential overdispersion[1].

Let $Y_{it}$ represent the total number of contacts for the individual $i$ observed at wave $t$. We assume that $Y_{it}$ follows a negative binomial distribution with mean $\mu_{it}$ and dispersion parameter $\phi$:

$$Y_{it} | \, ui \sim NB(\mu_{it}, \phi)$$

with $ui \sim N(0, \sigma_u^2)$ being a random intercept that accounts for within-individual correlation across waves ($\sigma_u^2$). Note that, as observations are repeated across waves, capturing within-individual correlation is essential to correctly account for baseline differences among individuals.

In this framework, the conditional expected value $E(Y_{it}|ui) = \mu_{it}$ is modelled as:

$$log(\mu_{it}) = \beta \cdot X_i + \gamma \cdot Z_{it} + \delta \cdot Wave_t + \eta \cdot (Z_{it} \times Wave_t) + ui$$

where:

- $X_i$ is a vector of individual-level covariates that are constant across waves (e.g., age, sex, and occupation).
- $\beta$ is the vector of fixed effect coefficients for these fixed covariates.
- $Z_{it}$ is a vector of time-varying covariates that may change across waves (e.g. attendance at work/school, occupation, whether the response was provided on Sunday, or whether the respondent recently tested positive to COVID-19).
- $\gamma$ is the vector of fixed effect coefficients for these time-varying covariates.
- $Wave_t$ is an indicator for wave, with coefficient $\delta$, capturing systematic differences between waves.
- $Z_{it} \times Wave_t$ represents the interaction terms between the wave indicator and each time-varying covariate $Z_{it}$.
- $\eta$ is the vector of coefficients for these interaction terms, allowing us to test whether the effect of time-varying covariates is mediated by the time of the survey.
- $ui \sim N(0, \sigma_u^2)$ is the random intercept term.

This mathematical structure is referred to as Generalised Linear Mixed Effects Model (GLMM) with a negative binomial outcome distribution and a log link function.

Given the extensive set of 35 covariates (see Figure SI5 for a full list of covariates), we first assessed each variable's association with the outcome using separate regression models. The model formulation reported above is general and can apply to different set of covariates. In this case, additionally to the specific variable screened, each separate model included a "wave" term and an interaction with the covariate to evaluate potential time-varying effects.

All covariates showing significant associations were included in the final multivariate negative binomial regression model presented in the main manuscript. The same covariates were used to model the different outcome variables discussed as robustness checks in Section SI2.1 and SI2.3.

## 1.4 Computation of contact matrices

We constructed age-specific matrices of average total contacts for the Italian population, stratified by 15 age groups (0-4, 5-9, 10–14, 15-19, 20-24, 25-29, 30-34, 35-39, 40-44, 45-49, 50-54, 55-59, 60-64, 65-69, and 70+ years old). Matrices were constructed by sampling 1000-times with replacement sets of social contact survey responses of the size of the cleaned survey responses. Then, matrices were normalised to account for reciprocal social contacts using national census population data aggregated by the same age groups (see below). Reciprocity-adjusted contact matrices resulting from each independent sampling step were pooled together to provide age-stratified social contact average estimates and confidence intervals.

### Reciprocity adjustment

The idea behind reciprocal normalisation is to account for differences between the sample age distribution and the age distribution of the population under study. This is done considering the probability of an individual being included in the sample. The procedure adopted in this work follows the standard procedure described in full in Trentini et al.[2] and summarized as follows.

Let $P_a$ denote the number of participants in the $a$-th age group and let $c_{a,\tilde{a}}(i)$ denote the number of contacts a specific study participant $i$ of age within $a$ had with individuals of the $\tilde{a}$ age group. The total number of contacts $T_{a,\tilde{a}}$ that all study participants of age $a$ have with individuals of age $\tilde{a}$ can be computed as

$$T_{a,\tilde{a}} = \sum_{i=1}^{P_a} c_{a,\tilde{a}}(i)$$

The average contacts an individual of age group $a$ has with individuals of age within $\tilde{a}$ can be approximated by the average contacts that a participant of age $a$ with individuals of age $\tilde{a}$ as follows:

$$C_{a,\tilde{a}} = \frac{T_{a,\tilde{a}}}{P_a}$$

In principle, $T_{a,\tilde{a}}$ can be different from $T_{\tilde{a},a}$. However, in the case of social contacts, if we were to sample the entire Italian population this should not be the case as all reported contacts should be reciprocal. To correct matrices drawn from only a partial sampling of the population we can, however, reconstruct reciprocity by considering the probability of an individual being included in the sample and correcting it using the total number of contacts that all study participants of age $a$ have with individuals of age $\tilde{a}$ as a weighted average of the total contacts reported by participants of these two ages as follows. This can be expressed in a formula as:

$$T_{a,\tilde{a}}^{corrected} = \frac{P_a N_a C_{a,\tilde{a}} + P_{\tilde{a}} N_{\tilde{a}} C_{\tilde{a},a}}{P_a + P_{\tilde{a}}}$$

where $N_a$ is the population within the age group $a$ in Italy as reported by the national statistical office.

The adjusted average contacts an individual of age $a$ has with individuals of age $\tilde{a}$ was finally computed as

$$C_{a,\tilde{a}}^{corrected} = \frac{T_{a,\tilde{a}}^{corrected}}{N_a}$$

**Bootstrapping contact matrices**

All the matrices presented in the main manuscript are computed by performing a 1000-fold bootstrapping of responses from both waves. The average number of contacts is computed as the average across the reciprocity-adjusted matrices originated from independent samples of survey responses. An average contact matrix is computed as the average across all bootstrapped samples and age groups: $\bar{C}_{a,\tilde{a}}^{corrected} = \frac{1}{Q}\sum_{i=1}^{Q} C_{a,\tilde{a}}^{corrected}$, with Q being the number of bootstrapped samples (1000).

## 1.5 Effect of school and work attendance on the transmissibility of a generic respiratory virus

We assessed the effect of different scenarios of distance learning and work-from-home on the reduction of the basic reproduction number $R_0$ of a novel respiratory virus spreading in a fully susceptible population in the absence of other interventions.

### Baseline scenario

In the baseline scenario (s=0 in equations reported in the main text), all students are assumed to attend school in presence and the in-person workforce is set at pre-pandemic levels. To define $N_i^P(s=0)$ and $N_i^{NP}(s=0)$, we considered the Italian population by age as reported by Eurostat for the year 2022 and stratify it into 15 age groups, $N_i$ (0-4 years, 5-9, years, ..., 65-69 years, 70+ years ). We estimated the proportion of individuals of age-group $i$ enrolled to education level $k$, denoted by $\sigma_i^k$, based on data reported by Eurostat for the year 2021[4]. Specifically, we considered 5 different education levels:

1. early childhood education (International Standard Classification of education, ISCED 0[5]);
2. primary education (ISCED 1[5]);
3. lower secondary education (ISCED 2[5]);
4. upper secondary education (ISCED 3[5]);
5. tertiary education (ISCED 5-8[5]).

Students enrolled to ISCED level 4 were neglected as, in Italy, they represent less than 0.03% of Italian students[4]. We further considered the age-specific proportion of individuals of age-group $i$ working in-presence before the COVID-19 pandemic ($\omega_i^{\text{pre-pandemic}}$), as estimated by the Italian Workers Compensation Authority and reported in Marziano et al.[6]. The number of individuals of age $i$ attending schools or work in-person in the baseline scenario (s=0) is computed as:

$$N_i^P(s=0)= \left(\sum_{k=1}^{k=5} \sigma_i^k + \omega_i^{\text{pre-pandemic}}\right) N_i$$

and the number of individuals not attending is given by:

$$N_i^{NP}(s=0) = N_i - N_i^P(s=0).$$

### Alternative scenarios

We then considered a set of alternative scenarios combining the physical closure of different education levels, starting from tertiary education and progressively including lower levels, with decreasing levels of in-person work attendance.

Specifically, we considered the following levels of school closure (i.e., distance learning):

- no education level closed;
- only tertiary education closed (ISCED 5-8);

- tertiary and upper secondary education closed (ISCED 5-8 and ISCED 3);
- tertiary; upper secondary; lower secondary education closed (ISCED 5-8 and ISCED 3, ISCED 2);
- tertiary; upper secondary; lower secondary, primary closed (ISCED 5-8 and ISCED 3, ISCED 2, ISCED 1);
- All education levels closed, i.e. tertiary; upper secondary; lower secondary, primary and early childhood (ISCED 5-8 and ISCED 3, ISCED 2, ISCED 1, ISCED 0).

We also considered three levels of in-person work attendance:

- "Pre-pandemic", i.e. the age-specific proportion of in-person work attendance is set at pre-pandemic levels;
- "Sustainable", reflecting the proportion of in-person workers observed in Italy after the reopening of economic activities following the 2020 COVID-19 lockdown on May 18, 2020;
- "Minimum", where only essential workers are allowed to work in person, as during the COVID-19 lockdown.

The age-specific proportions of in-person workers associated to the "Sustainable" and "Minimum" scenarios ($\omega_i^{\text{sustainable}}$ and $\omega_i^{\text{minimum}}$) were also estimated by the Italian Workers Compensation Authority and are available from Marziano et al.[6] and reported in Figure 3B in the main text.

The resulting 18 scenarios (including the baseline) are reported in Table SI2.

**Table SI2:** *Description of intervention scenarios considered.*

| Level of school closure | Level of in-person work | | |
|---|---|---|---|
| | Pre-pandemic | Sustainable *(no closure of economic activities)* | Minimum *(only essential activities)* |
| **No level closed** | Scenario 0 *(Baseline)* | Scenario 1 | Scenario 2 |
| **ISCED 5-8** *(tertiary)* | Scenario 3 | Scenario 4 | Scenario 5 |
| **ISCED 3-8** *(tertiary and upper secondary)* | Scenario 6 | Scenario 7 | Scenario 8 |
| **ISCED 2-8** *(tertiary; upper secondary; lower secondary)* | Scenario 9 | Scenario 10 | Scenario 11 |
| **ISCED 1-8** *(tertiary; upper secondary; lower secondary; primary)* | Scenario 12 | Scenario 13 | Scenario 14 |
| **All levels closed** *(from early childhood to tertiary)* | Scenario 15 | Scenario 16 | Scenario 17 |

## 2. Additional figures and results

### 2.1 Main analysis

#### Comparing the data with previous studies: POLYMOD[1] and CoMix[7]

The data presented in the main manuscript complement previous social contact data, providing updated information collected during a period when COVID-19 measures in Italy (both at the governmental and individual level) were receding (wave 1) or had been removed altogether (wave 2). In particular, the data from the second wave were collected only one month before the WHO declared the end of the COVID-19 epidemic as a public health emergency of international concern[8]. We compared data with two previous social contact study in Italy: POLYMOD[9], a paper-based survey conducted in April 2006; and CoMix - Italy, a series of online surveys conducted at different times during the pandemics. For CoMix, we selected Wave A6 and A7 as they took place in 2021 roughly in the same months as our study (see Epipose Project Italy Report). As A6 and A7 only included adult respondents, we similarly pooled data about the underage population from waves C1 and C2 (which were survey waves focused on the underage population only). During this period, the SARS-CoV-2 Alpha variant was dominant in Italy, COVID-19 vaccination was being rolled out and strict COVID-19 control measures were in place. Tables SI3 and Table SI4 report the number of participants and average reported contacts for the three studies, stratifying the information by different groups of respondents. Data for comparison were obtained through the *socialmixr* project[9].

**Table SI3:** *Survey respondents and average contacts by wave - comparing Polymod, CoMix, and our study.* The table reports the sample size and average number of social contacts stratified by different respondent characteristics (95% confidence intervals are reported in parenthesis): age, sex, household size, and day of the week.

| | Category | Polymod April 2006 count | Polymod April 2006 mean | CoMix February-April 2021 (A6, A7, C1, C2) count | CoMix February-April 2021 (A6, A7, C1, C2) mean | This study March 2022 count | This study March 2022 mean | This study March-April 2023 count | This study March-April 2023 mean |
|---|---|---|---|---|---|---|---|---|---|
| All | **All** | 846 | 19.8 (19 - 20.6) | 1883 | 3.6 (3.3 - 3.9) | 2474 | 7.3 (6.9 - 7.7) | 2505 | 7.8 (7.3 - 8.2) |
| Age | **0-4** | 77 | 14.9 (12.7 - 17.1) | 433 | 6.2 (5.3 - 7.1) | 76 | 9 (7.2 - 10.9) | 54 | 12 (8.9 - 15.5) |
| | **5-17** | 224 | 24.9 (23.2 - 26.5) | 295 | 2.6 (2.3 - 3) | 233 | 12.8 (11.4 - 14.3) | 156 | 14.4 (12.4 - 16.5) |
| | **18-29** | 112 | 20.7 (18.5 - 23) | 161 | 5.7 (4.7 - 6.8) | 208 | 10.6 (8.7 - 12.8) | 190 | 10.9 (9 - 13.1) |
| | **30-39** | 104 | 18.4 (16.5 - 20.4) | 229 | 2.8 (2.2 - 3.6) | 376 | 8 (6.8 - 9.3) | 357 | 8.8 (7.7 - 10.1) |
| | **40-49** | 119 | 19.3 (17.1 - 21.5) | 198 | 2.4 (2 - 3.1) | 381 | 6 (5.1 - 7.2) | 339 | 6.7 (5.8 - 7.8) |
| | **50-59** | 105 | 18.9 (16.6 - 21.3) | 151 | 2.8 (1.9 - 4.4) | 432 | 6.3 (5.4 - 7.2) | 506 | 7.4 (6.5 - 8.4) |
| | **60+** | 98 | 13.7 (11.8 - 15.8) | 407 | 1.9 (1.8 - 2.1) | 768 | 5.5 (4.9 - 6.1) | 903 | 5.9 (5.3 - 6.6) |
| | **NA** | 7 | 23.4 (16.9 - 29) | 9 | 6 (1.9 - 13) | - | - | - | - |
| Sex | **Female** | 438 | 19.3 (18.1 - 20.4) | 1176 | 4.3 (3.8 - 4.7) | 1226 | 7.1 (6.6 - 7.7) | 1233 | 7.3 (6.8 - 7.9) |
| | **Male** | 401 | 20.3 (19.1 - 21.5) | 703 | 2.4 (2.2 - 2.6) | 1248 | 7.5 (6.9 - 8.1) | 1272 | 8.2 (7.6 - 8.9) |
| | **NA** | 7 | 23.4 (17 - 29.1) | 4 | 13 (7 - 22.8) | - | - | - | - |
| Household Size | **1** | 48 | 15 (11.8 - 18.4) | 171 | 0.9 (0.6 - 1.1) | 357 | 5.2 (4.5 - 6) | 355 | 5.9 (4.9 - 7) |
| | **2** | 110 | 17.2 (15.1 - 19.4) | 439 | 2 (1.6 - 2.6) | 745 | 5.7 (5 - 6.5) | 844 | 6.5 (5.7 - 7.2) |
| | **3** | 227 | 18.4 (16.9 - 20) | 577 | 3.6 (3.1 - 4.1) | 732 | 8.2 (7.3 - 9.1) | 677 | 8.5 (7.6 - 9.4) |
| | **4** | 333 | 21.7 (20.5 - 23) | 555 | 5.1 (4.5 - 5.8) | 502 | 9.1 (8.2 - 10.1) | 492 | 9.6 (8.6 - 10.6) |
| | **5** | 98 | 21.1 (18.4 - 23.9) | 107 | 5.4 (4.1 - 7.3) | 114 | 9.7 (8.1 - 11.7) | 104 | 10.3 (8.4 - 12.6) |
| | **6+** | 30 | 21 (16.7 - 25.5) | 33 | 6.7 (5.2 - 8.8) | 24 | 13.9 (8.2 - 22.3) | 33 | 13.5 (9.6 - 17.9) |
| | **NA** | - | - | 1 | 2 (1 - 2) | - | - | - | - |
| Day | **weekday** | 620 | 21.1 (20.2 - 22.1) | 1565 | 3.8 (3.5 - 4.1) | 1827 | 7.4 (6.9 - 7.9) | 1757 | 8.3 (7.8 - 8.9) |
| | **weekend** | 226 | 16 (14.6 - 17.5) | 318 | 2.6 (2.1 - 3.4) | 647 | 7.2 (6.3 - 8.2) | 748 | 6.5 (5.8 - 7.3) |

**Table SI4:** *Survey average contacts by wave and contact setting - comparing Polymod, CoMix, and our study*. The table reports the sample size and average number of social contacts in different location settings (95% confidence intervals are reported in parenthesis): home, work, school, transportation means, leisure places, and other locations.

| Setting | Polymod | | CoMix | | This study | | | |
|---|---|---|---|---|---|---|---|---|
| | April 2006 | | February-April 2021 (A6, A7, C1, C2) | | March 2022 wave | | March 2023 wave | |
| | count | mean | count | mean | count | mean | count | mean |
| **home** | 781 | 3.8 (3.7 - 4) | 1492 | 1.8 (1.7 - 1.9) | 2206 | 4.8 (4.5 - 5.2) | 2268 | 4.9 (4.6 - 5.2) |
| **work** | 258 | 3.4 (2.9 - 3.9) | 128 | 0.2 (0.1 - 0.4) | 375 | 0.7 (0.6 - 0.9) | 403 | 0.9 (0.8 - 1.1) |
| **school** | 323 | 5.3 (4.7 - 5.9) | 158 | 0.8 (0.6 - 1) | 118 | 0.4 (0.3 - 0.5) | 98 | 0.4 (0.3 - 0.5) |
| **leisure** | 527 | 3.7 (3.3 - 4) | 15 | 0 (0 - 0) | 394 | 0.5 (0.5 - 0.6) | 474 | 0.6 (0.5 - 0.7) |
| **transport** | 162 | 0.5 (0.4 - 0.7) | 23 | 0 (0 - 0) | 121 | 0.1 (0.1 - 0.2) | 132 | 0.1 (0.1 - 0.1) |
| **other** | 490 | 3 (2.7 - 3.4) | 469 | 0.7 (0.6 - 0.9) | 648 | 0.7 (0.6 - 0.8) | 724 | 0.9 (0.8 - 1) |

## Modelling the determinants of social contacts: covariates description

**Table SI5:** *Description of the covariates used in the statistical modelling framework.* The table reports the name, type, description, and possible values of all the covariates used in the negative binomial regression.

| NAME | TYPE | DESCRIPTION | VALUES |
|---|---|---|---|
| **REGION_GROUPED_IT** | Categorical | Italian macro-region of residence of the respondent (ISTAT categorization) | • North-West<br>• North-East<br>• Center<br>• South<br>• Islands |
| **RESPONDENT_GENDER** | Binary | Gender of the respondent | • Male<br>• Female |
| **EDUCATION** | Categorical | Over 17: Educational level achieved by the respondent (ISCED categorization)<br>Under 17: Educational level achieved by the respondent's parent (ISCED categorization) | • Lower secondary or below<br>• Upper secondary<br>• Tertiary or above |
| **OCCUPATION_AGG** | Categorical | Over 16: Respondent's job category<br>Under 16: Job category of the respondent's parent | • Employed<br>• Home/family<br>• Student<br>• Retired<br>• Inactive |
| **D_OCCUPATION** | Binary | Over 16: Respondent's job category<br>Under 16: Job category of the respondent's parent | • Employed/student<br>• Other (not working) |
| **OCCUPATION_AGG2** | Categorical | Over 16: Respondent's job category<br>Under 16: Job category of the respondent's parent | • Employed<br>• Student<br>• Other (not working) |
| **SUNDAY** | Binary | Contact diary refers to Sunday | • Yes<br>• No |
| **WEEKEND** | Binary | Contact diary refers either to Saturday or Sunday | • Yes<br>• No |
| **WORKED_IN_PRESENCE** | Binary | Over 16: The respondent worked in-presence on the day prior to survey filling | • Yes<br>• No (remote work, no work) |
| **PRESENCE_SCHOOL** | Categorical | Under 16, students over 16: teaching method followed on the day prior to the survey filling | • In person<br>• Remotely<br>• Did not attend |
| **ATTEND_IN_PRESENCE** | Binary | The respondent worked or followed class at school in-presence the day prior to the survey filling | • Yes<br>• No |
| **D_CHILDREN** | Binary | The respondent lives with at least one minor in their household | • Yes<br>• No |
| **D_SENIOR65** | Binary | The respondent lives with at least one person aged 65 or more in their household | • Yes<br>• No |
| **D_SENIOR70** | Binary | The respondent lives with at least one person aged 70 or more in their household | • Yes<br>• No |
| **D_SENIOR75** | Binary | The respondent lives with at least one person aged 75 or more in their household | • Yes<br>• No |
| **D_SENIOR80** | Binary | The respondent lives with at least one person aged 80 or more in their household | • Yes<br>• No |
| **D_HH_SIZE** | Binary | The respondent has at least a cohabitant | • Yes<br>• No |
| **HH_SIZE** | Categorical | Household size (respondent included) | • 1<br>• 2<br>• 3<br>• 4<br>• 5+ |
| **HH_SIZE_DET** | Numeric | Household size (respondent included) | • [1-8] |
| **CHRONIC_COMORB_SELF2** | Binary | The respondent has at least a comorbidity known to be associated with severe COVID-19 outcomes[1] | • Yes<br>• No |
| **CHRONIC_COMORB_COHAB2** | Binary | At least one of the respondent's cohabitant has at least a comorbidity known to be associated with severe COVID-19 outcomes[1] | • Yes<br>• No |
| **COVID** | Binary | The respondent has been officially diagnosed with COVID-19 at least once | • Yes<br>• No |
| **TIME_SINCE_COVID** | Categorical | Time passed since last officially diagnosed COVID-19 infection | • Had COVID-19 – less than 120 days ago<br>• Had COVID-19 – more than 120 days ago<br>• Had COVID-19 – but reported no date<br>• Didn't have COVID-19 |
| **D_TIME_SINCE_COVID** | Binary | The respondent has been officially diagnosed with COVID-19 in the 4 months prior to the survey filling | • Yes<br>• No |
| **INCOME_1000** | Categorical | Respondent's family income | • <1000 eur<br>• 1000-1999 eur<br>• 2000-2999 eur<br>• 3000-3999 eur<br>• 4000-4999 eur<br>• >5000 eur<br>• Didn't answer |
| **INCOME** | Categorical | Respondent's family income | • <1000 eur<br>• 1000-1499 eur<br>• 1500-1999 eur<br>• 2000-2499 eur<br>• 2499-2999 eur<br>• 3000-3499 eur<br>• 3500-3999 eur<br>• 4000-4499 eur<br>• 4500-4999 eur<br>• >5000 eur<br>• Didn't answer |
| **INCOME_THREECAT** | Categorical | Respondent's family income | • <1500 eur<br>• 1500-2999 eur<br>• >3000 eur<br>• Didn't answer |
| **KNOWS_SEVERE** | Binary | The respondent knows someone who was hospitalized/died due to COVID-19 | • Yes<br>• No |
| **AGE_GROUP_10** | Categorical | Age group of the respondent | • 0-17 y<br>• 18-29 y<br>• 30-39 y<br>• 40-49 y<br>• 50-59 y<br>• 60-69 y<br>• 70+ y |
| **AGE_GROUP_25** | Categorical | Age group of the respondent | • 0-17 y<br>• 18-44 y<br>• 45-69 y<br>• 70+ y |
| **WAVE** | Categorical | Survey wave | • 1<br>• 2 |
| **EPID** | - | Alphanumeric code that uniquely identifies respondents across survey waves | - |

---

[1] Alcohol or drug consumption abuse; Sickle cell disease or thalassemia; Neurological diseases; Pregnancy; AIDS/HIV infection; Chronic liver diseases; Chronic kidney diseases; Chronic lung diseases (e.g., COPD, asthma, cystic fibrosis, etc.).

## Modelling the determinants of social contacts for the underage population under scenario 1A assumptions

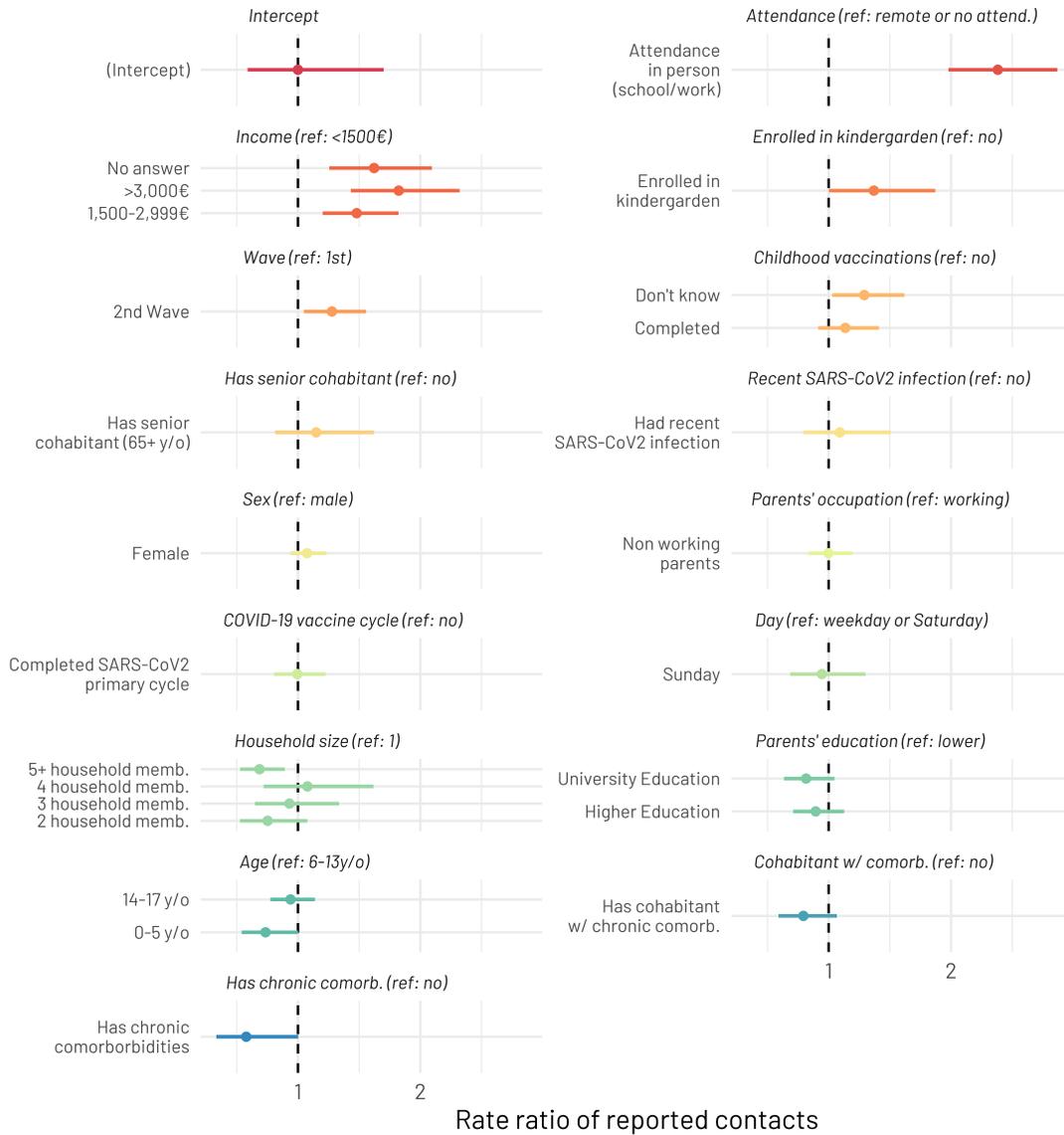

Rate ratio of reported contacts

**Figure SI4:** *Model results for the underage population under scenario 1A assumptions.* Each panel in the figure shows the risk ratio of reporting a larger number of direct and indirect social contacts in the underage population using prolonged proportional contacts for each of the model's covariates.

## Contact matrices by setting

This section reports the social contact matrices by setting for direct and indirect social under the assumptions describer as *scenario 1A* (see Section SI1.2 and Table SI1).

Figure SI5 shows the contact matrices for the entire survey population.

Figure SI6 and Figure SI7 respectively show the contact matrices for the survey respondents attending "in-person" work or school activities and for those not attending work or school "in-person".

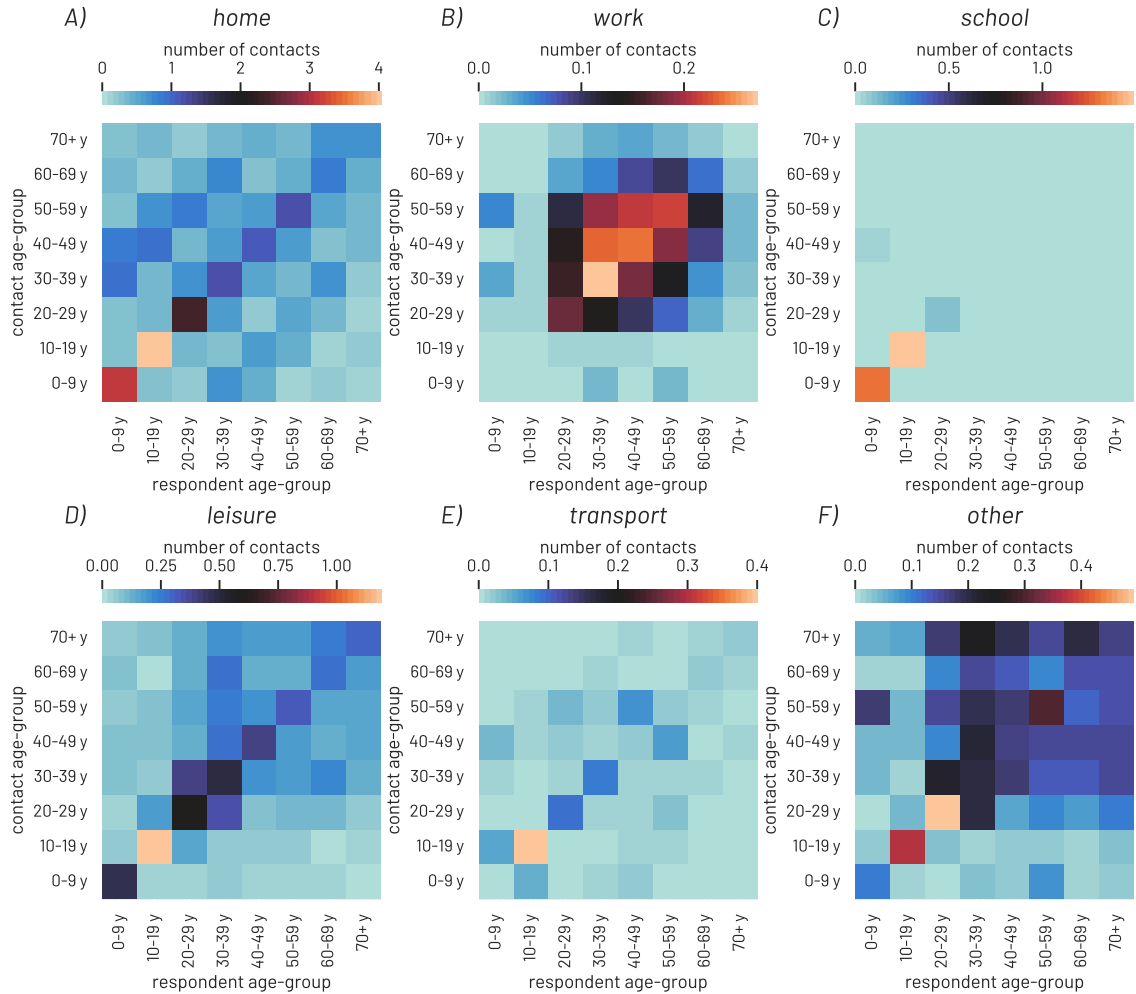

**Figure SI5:** *Contact matrices by setting under scenario 1A assumptions on social contacts.*

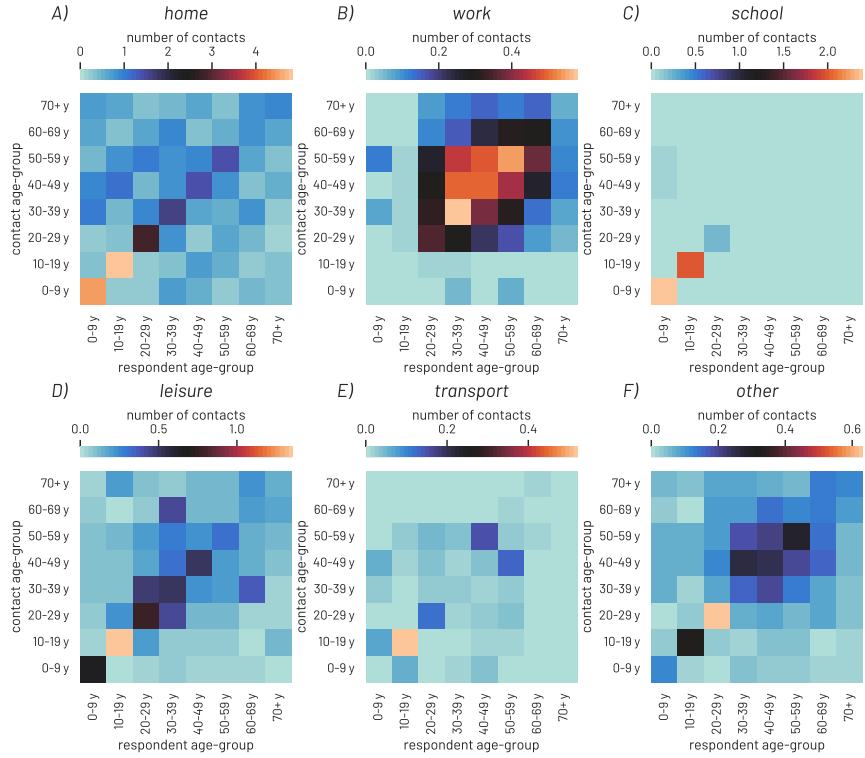

**Figure SI6:** *Contact matrices by setting of individuals working or attending school in-person under scenario 1A assumptions.*

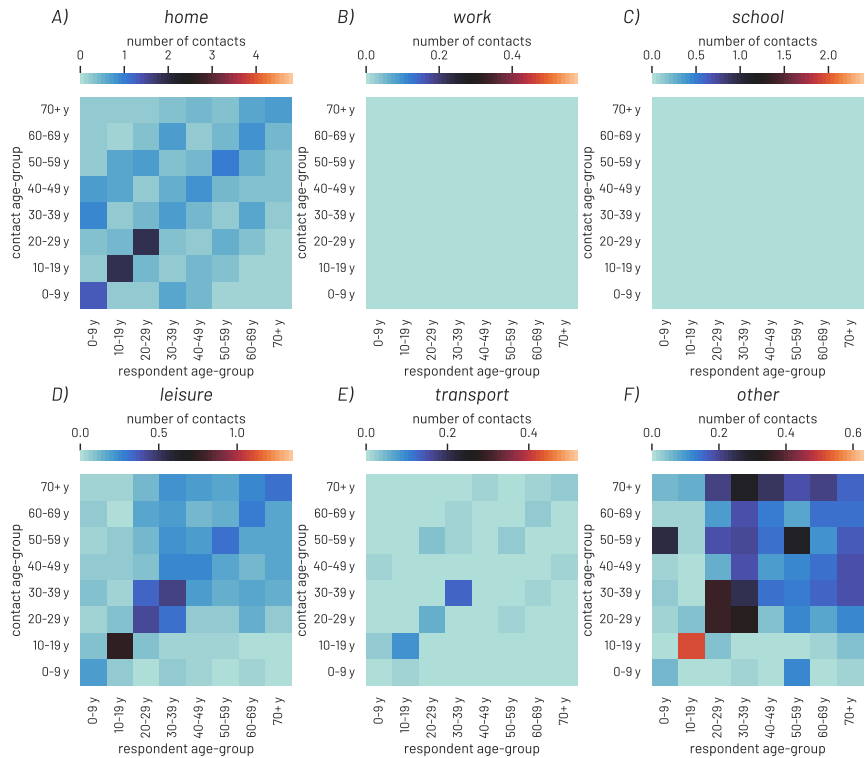

**Figure SI7:** *Contact matrices by setting for individuals not working in-person nor attending school in-person under scenario 1A assumptions.*

## 2.2 Additional details on direct and indirect social contacts

In this section, we separately report contact information stratified by respondent characteristics for direct and indirect contacts. The numbers reported in Table SI6 and SI7 show the average reported contacts for the two contact types. Table SI8 report the contacts by setting for direct, indirect and the minimum set union, i.e. the total social contacts as presented in the main manuscript, for the different settings.

**Table SI6:** *Survey average contacts by wave and contact type.* The table reports the average number of direct and indirect social contacts (and 95% C.I.) stratified by different respondent characteristics: age, sex, household income, and household size.

| Variables | March 2022 wave | | March 2023 wave | |
|---|---|---|---|---|
| | Direct Contacts (95% C.I.) | Indirect Contacts (95% C.I.) | Direct Contacts (95% C.I.) | Indirect Contacts (95% C.I.) |
| **Total** | 4.6 (4.4 - 4.8) | 5.7 (5.2 - 6.1) | 4.9 (4.7 - 5.1) | 6 (5.6 - 6.4) |
| **Age group** | | | | |
| 0-9 | 6.8 (5.8 - 7.8) | 9.7 (8.2 - 11.4) | 7.2 (6.2 - 8.3) | 10.6 (8.5 - 12.8) |
| 10-19 | 6.7 (5.9 - 7.6) | 11 (9.2 - 12.9) | 7 (5.8 - 8.3) | 13.6 (11 - 16.3) |
| 20-29 | 5.1 (4.6 - 5.7) | 8.8 (6.9 - 11) | 5.7 (5.1 - 6.3) | 9.1 (7 - 11.4) |
| 30-39 | 5.1 (4.7 - 5.5) | 6.3 (5.2 - 7.6) | 5.3 (4.9 - 5.8) | 7.1 (5.9 - 8.3) |
| 40-49 | 4.3 (4 - 4.7) | 4.5 (3.6 - 5.6) | 4.6 (4.3 - 5) | 5.2 (4.3 - 6.2) |
| 50-59 | 4.2 (3.8 - 4.5) | 4.6 (3.8 - 5.5) | 4.8 (4.4 - 5.2) | 5.8 (4.9 - 6.8) |
| 60-69 | 4.1 (3.7 - 4.5) | 4.7 (3.6 - 5.8) | 4.4 (4 - 4.8) | 4.4 (3.5 - 5.5) |
| 70+ | 3.4 (3.1 - 3.7) | 3 (2.4 - 3.7) | 4.2 (3.8 - 4.6) | 3.7 (3.1 - 4.4) |
| **Sex** | | | | |
| Male | 4.5 (4.3 - 4.7) | 5.7 (5.2 - 6.3) | 5 (4.8 - 5.3) | 6.3 (5.7 - 7) |
| Female | 4.7 (4.5 - 5) | 5.6 (5 - 6.2) | 4.8 (4.6 - 5.1) | 5.7 (5.2 - 6.2) |
| **Household income** | | | | |
| <1,500 euro | 3.7 (3.4 - 4) | 4.3 (3.6 - 5.2) | 3.9 (3.6 - 4.2) | 4.8 (4 - 5.7) |
| 1,500-2,999 euro | 4.8 (4.5 - 5) | 5.7 (5.1 - 6.5) | 5.1 (4.8 - 5.4) | 5.8 (5.3 - 6.4) |
| >3,000 euro | 5.8 (5.4 - 6.2) | 7 (6 - 8) | 6.2 (5.8 - 6.6) | 8.2 (7 - 9.4) |
| Doesn't answer | 4.1 (3.7 - 4.5) | 5.8 (4.8 - 6.8) | 4.3 (3.9 - 4.7) | 5.2 (4.3 - 6.2) |
| **Household size** | | | | |
| 1 | 3.4 (3 - 3.8) | 3.5 (2.8 - 4.3) | 3.5 (3.2 - 3.9) | 4.4 (3.5 - 5.5) |
| 2 | 3.6 (3.3 - 3.8) | 4.2 (3.5 - 4.9) | 3.9 (3.7 - 4.2) | 5 (4.2 - 5.7) |
| 3 | 4.8 (4.5 - 5.1) | 6.5 (5.7 - 7.5) | 5.3 (4.9 - 5.7) | 6.6 (5.8 - 7.5) |
| 4 | 5.9 (5.5 - 6.3) | 7.4 (6.5 - 8.4) | 6.4 (6 - 6.8) | 7.5 (6.6 - 8.5) |
| 5+ | 7.4 (6.7 - 8.2) | 8 (6.2 - 10.2) | 7.6 (6.7 - 8.6) | 8.4 (6.6 - 10.5) |

**Table SI7:** *Survey average contacts by wave and contact type.* The table reports the average number of direct and indirect social contacts (and 95% C.I.) stratified by different respondent characteristics, only for respondents aged 18 and older: educational attainment, occupation, working mode (e.g. remotely or in-presence), teaching method (e.g. distance or in-presence learning), COVID-19 vaccinal status, and COVID-19 verified infections.

| Variables | March 2022 wave | | March 2023 wave | |
|---|---|---|---|---|
| | **Direct Contacts (95% C.I.)** | **Indirect Contacts (95% C.I.)** | **Direct Contacts (95% C.I.)** | **Indirect Contacts (95% C.I.)** |
| **<u>Total</u>** | 4.3 (4.1 - 4.5) | 5 (4.6 - 5.4) | 4.7 (4.5 - 4.9) | 5.5 (5 - 5.9) |
| **<u>Educational attainment</u>** | | | | |
| Lower secondary or below | 3.5 (3.2 - 3.9) | 4.4 (3.4 - 5.6) | 3.8 (3.4 - 4.2) | 4.1 (3.2 - 5.2) |
| Upper secondary | 4.2 (4 - 4.4) | 4.5 (4 - 5.1) | 4.6 (4.4 - 4.8) | 5 (4.5 - 5.5) |
| Tertiary or above | 4.8 (4.5 - 5.1) | 6.1 (5.2 - 7) | 5.4 (5 - 5.7) | 6.9 (6 - 7.9) |
| **<u>Employment status</u>** | | | | |
| Employed (full or part-time) | 5 (4.8 - 5.3) | 5.8 (5.2 - 6.5) | 5.4 (5.1 - 5.7) | 6.7 (6.1 - 7.4) |
| Looking after home/family | 3.5 (3.1 - 3.9) | 2.7 (2.2 - 3.2) | 4.1 (3.7 - 4.6) | 4.5 (3.5 - 5.6) |
| Student (full- or part-time) | 4.8 (4.1 - 5.6) | 10.6 (7.4 - 14.4) | 5.4 (4.5 - 6.3) | 10.3 (6.6 - 14.6) |
| Retired | 3.4 (3.1 - 3.7) | 3.3 (2.6 - 4) | 4.1 (3.7 - 4.4) | 3.5 (2.9 - 4.1) |
| Inactive | 3.3 (3 - 3.7) | 4.5 (3.1 - 6) | 3.2 (2.9 - 3.6) | 3.7 (2.8 - 4.6) |
| **<u>Working mode (only employed respondents during weekdays)</u>** | | | | |
| Remote or did not attend | 3.5 (3.1 - 3.9) | 3.5 (2.7 - 4.3) | 4 (3.5 - 4.5) | 4.1 (3.2 - 5.2) |
| In-presence | 5.9 (5.5 - 6.2) | 6.4 (5.6 - 7.3) | 6.4 (6 - 6.8) | 8 (7 - 9.1) |
| **<u>Teaching method (only student respondents during weekdays)</u>** | | | | |
| In-presence | 5.5 (4.1 - 6.9) | 13.6 (8.6 - 19.2) | 5.5 (4.2 - 6.8) | 15 (7.7 - 23.6) |
| Distance learning | 3.4 (2.5 - 4.4) | 6.2 (3.1 - 10.3) | - | - |
| Did not attend | 4.7 (3 - 6.7) | 8.1 (2.4 - 15) | 4.9 (3.5 - 6.5) | 5.1 (2.9 - 8.4) |
| **<u>COVID-19 vaccination status</u>** | | | | |
| None | 3.8 (3.2 - 4.5) | 4 (2.8 - 5.5) | 3.6 (3.1 - 4.1) | 3.2 (2.5 - 4.1) |
| 1 dose | 3.5 (1.9 - 5.2) | 2.1 (0.8 - 3.6) | 5.3 (3.4 - 7.7) | 4.8 (2.5 - 7.7) |
| 2 doses | 4 (3.6 - 4.5) | 5.3 (3.9 - 7.1) | 4.8 (4.3 - 5.4) | 5.8 (4.5 - 7.3) |
| 3 doses or more | 4.4 (4.2 - 4.6) | 5.1 (4.6 - 5.6) | 4.8 (4.6 - 5) | 5.6 (5.1 - 6.1) |
| Vaccine Exemption | 2.1 (1.3 - 2.9) | 5.4 (0.9 - 13.6) | 8.3 (4 - 12.9) | 11.3 (2.2 - 28.4) |
| **<u>COVID-19 previous infection</u>** | | | | |
| Yes | 4.8 (4.4 - 5.2) | 5.8 (4.6 - 7) | 4.9 (4.7 - 5.2) | 6 (5.4 - 6.7) |
| No | 4.2 (4 - 4.4) | 4.8 (4.4 - 5.3) | 4.5 (4.3 - 4.8) | 4.9 (4.4 - 5.5) |

**Table SI8:** *Survey average contacts by wave and contact setting.* The table reports the average number of direct, indirect, and total social contacts (and 95% C.I.) in different location settings: home, work, school, leisure places, transportation means, and other locations.

| Variable | March 2022 wave | | | March 2023 wave | | |
|---|---|---|---|---|---|---|
| **Contacts** | **Direct (95% C.I.)** | **Indirect (95% C.I.)** | **Total (95% C.I.)** | **Direct (95% C.I.)** | **Indirect (95% C.I.)** | **Total (95% C.I.)** |
| <u>**Total**</u> | 4.6 (4.4 - 4.8) | 5.7 (5.2 - 6.1) | 7.3 (6.9 - 7.8) | 4.9 (4.7 - 5.1) | 6 (5.6 - 6.4) | 7.8 (7.3 - 8.2) |
| <u>**Contact setting**</u> | | | | | | |
| Home | 3 (2.9 - 3.1) | 3.8 (3.5 - 4.2) | 4.8 (4.5 - 5.2) | 3.1 (3 - 3.3) | 3.8 (3.5 - 4.1) | 4.9 (4.6 - 5.2) |
| Work | 0.5 (0.4 - 0.6) | 0.6 (0.5 - 0.7) | 0.7 (0.6 - 0.9) | 0.6 (0.5 - 0.6) | 0.8 (0.6 - 0.9) | 0.9 (0.8 - 1.1) |
| School | 0.2 (0.2 - 0.3) | 0.4 (0.3 - 0.4) | 0.4 (0.3 - 0.5) | 0.2 (0.1 - 0.2) | 0.3 (0.2 - 0.4) | 0.4 (0.3 - 0.5) |
| Leisure | 0.4 (0.3 - 0.4) | 0.3 (0.3 - 0.4) | 0.5 (0.5 - 0.6) | 0.4 (0.4 - 0.5) | 0.4 (0.4 - 0.5) | 0.6 (0.5 - 0.7) |
| Transport | 0.1 (0.1 - 0.1) | 0.1 (0.1 - 0.2) | 0.1 (0.1 - 0.2) | 0.1 (0.1 - 0.1) | 0.1 (0.1 - 0.1) | 0.1 (0.1 - 0.1) |
| Other | 0.5 (0.4 - 0.5) | 0.4 (0.4 - 0.5) | 0.7 (0.6 - 0.8) | 0.6 (0.5 - 0.6) | 0.6 (0.5 - 0.7) | 0.9 (0.8 - 1) |

## 2.3 Sensitivity analysis on direct contacts only

This section is devoted to presenting the same analyses presented in the main manuscript but focusing exclusively on direct contacts only, i.e. only on contacts reported by respondents within the social contact diary section of the survey.

### Modelling the determinants of direct social contacts

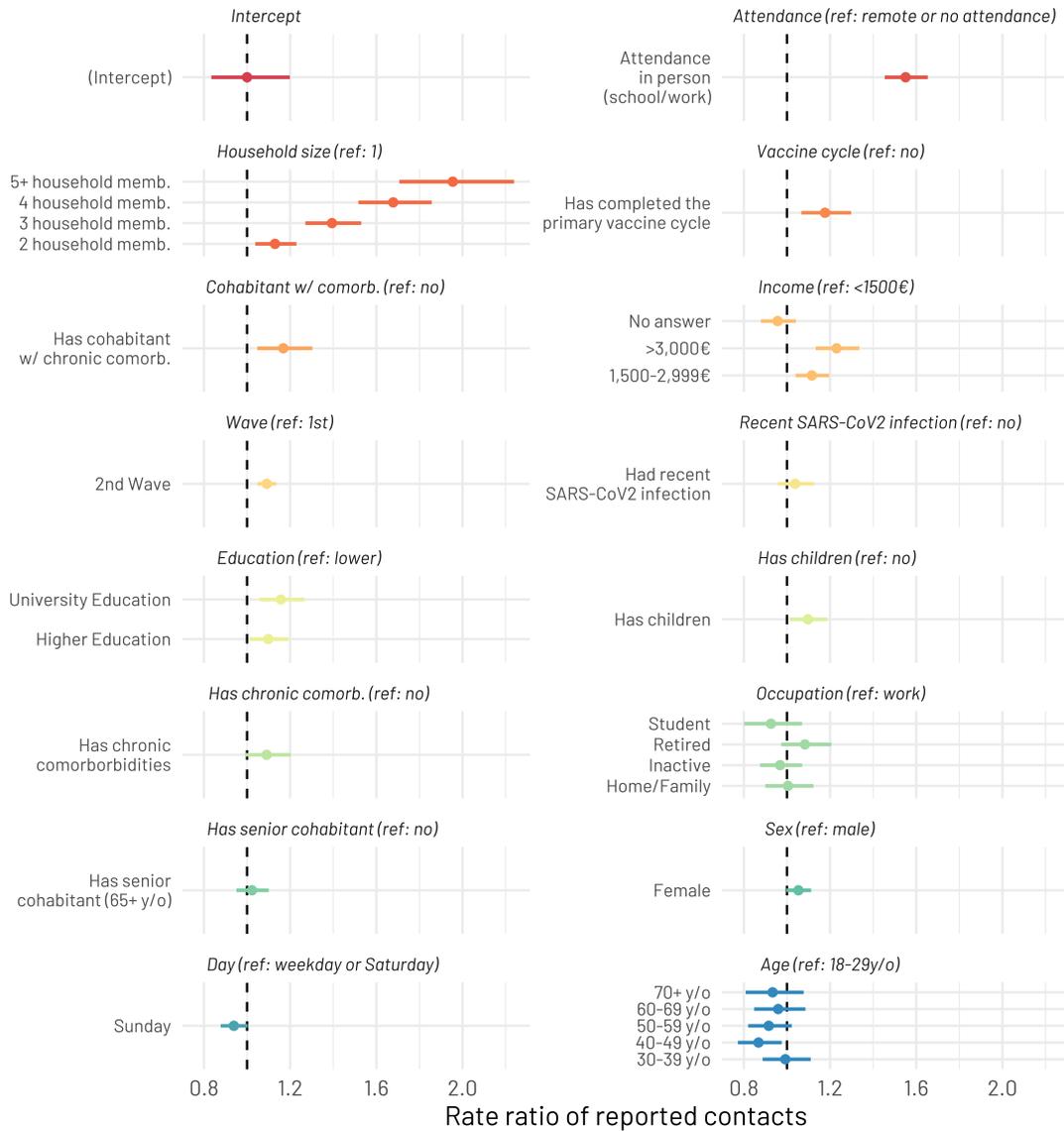

**Figure SI8:** *Model results for the direct social contacts of the adult population.* Each panel in the figure shows the risk ratio of reporting a larger number of direct social contacts in the adult population for each of the model's covariates.

## Social contacts matrices for direct contacts

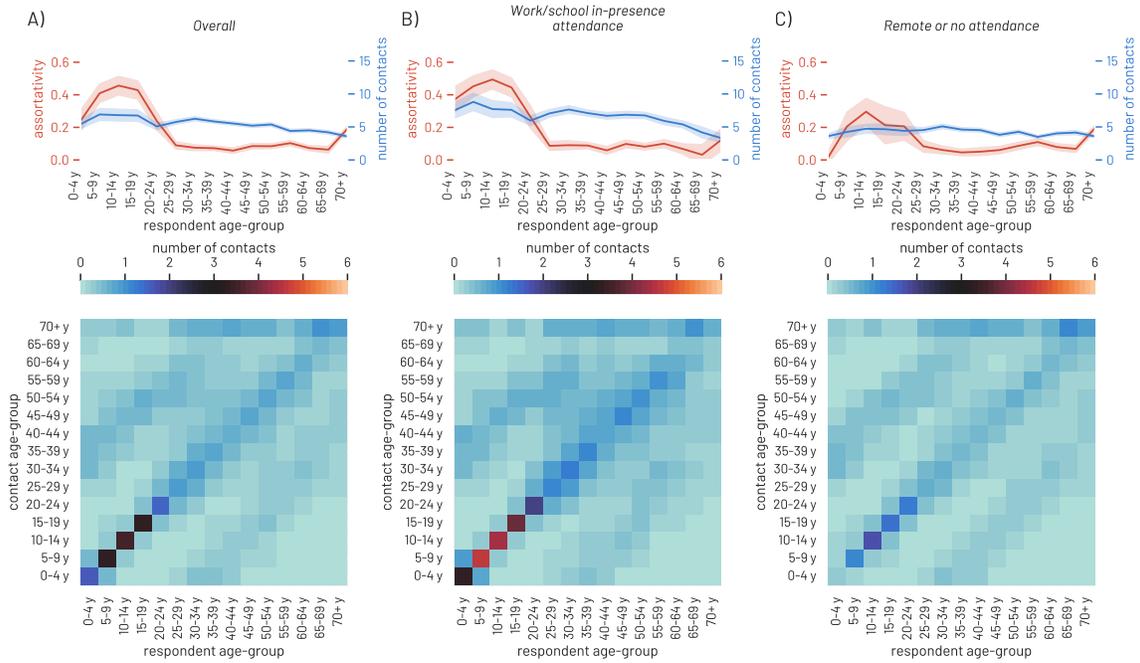

**Figure SI9:** *Contact matrices for direct-only social contacts.* A) Overall direct contacts, B) Contacts of respondents who attended in-person work or school activities, C) Contacts of respondents who did not attend work/school in-person.

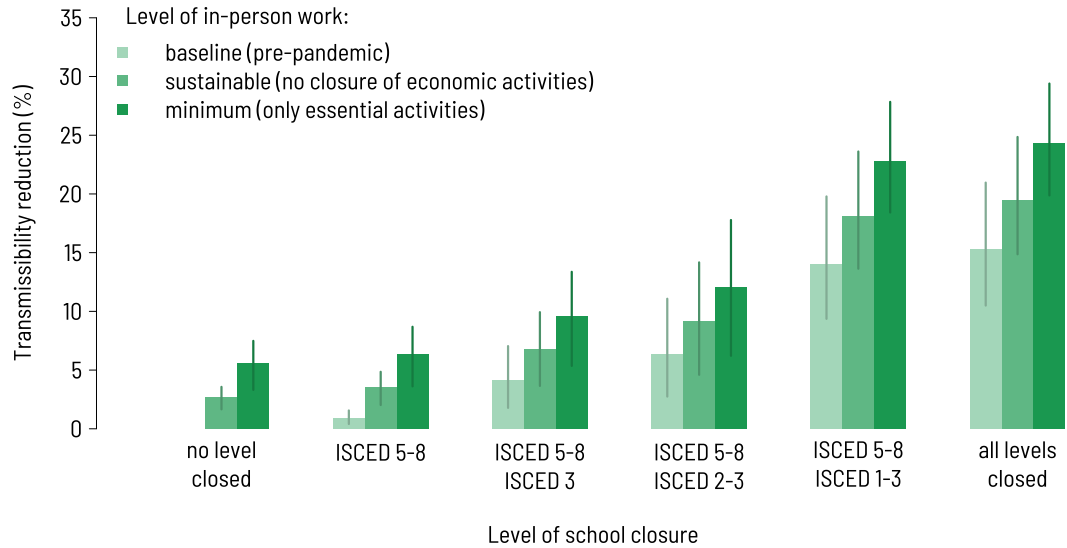

**Figure SI10:** *Effectiveness of in-person school and work closures in reducing viral transmissibility in the sensitivity analysis considering direct contacts only.* The Figure shows the relative reduction in transmissibility associated with intervention scenarios combining different levels of school closures and in-person work, compared to a baseline scenario where the proportion of in-person workers is set at pre-pandemic levels[6] and students across all education levels attend in person.

## 2.4 Sensitivity analysis for social contacts under scenario 1B

This section is devoted to presenting the same analyses discussed in the main manuscript but adopting a different indirect social contacts enhancement strategy: the "dominant setting". The sensitivity analysis discussed here corresponds to scenario 1B discussed in Section SI1.2.

### Social contacts matrices

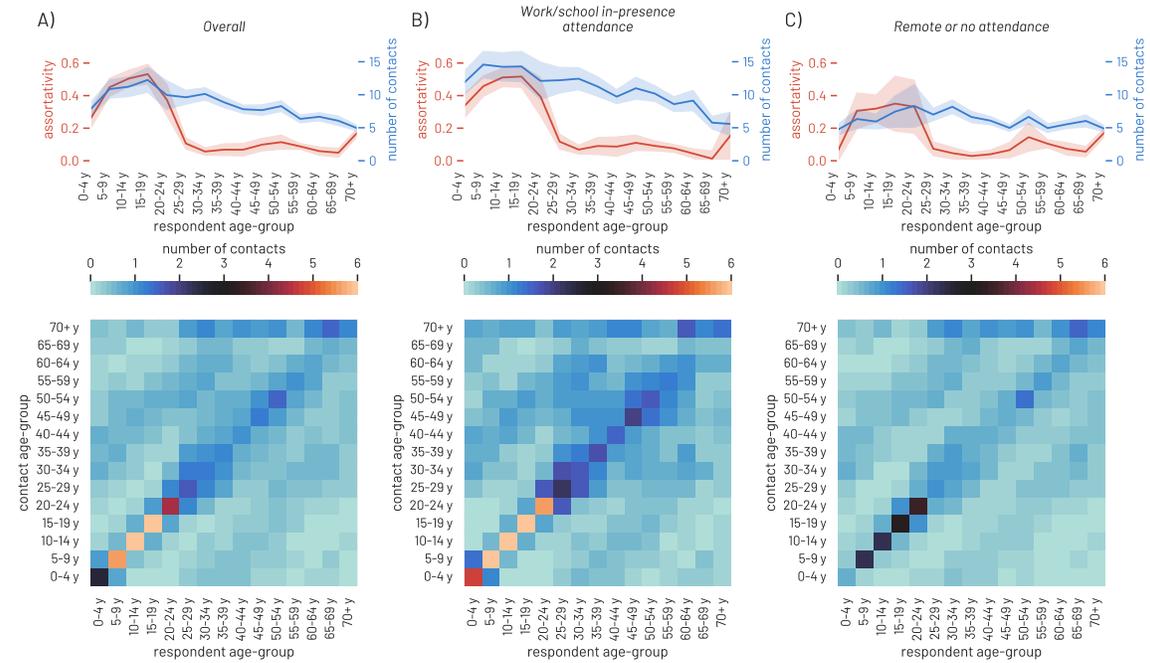

**Figure SI11:** *Contact matrices for social contacts under scenario 1B assumptions.* A) Overall direct and indirect B1 social contacts, B) Contacts of respondents who attended in-person work or school activities, C) Contacts of respondents who did not attend work or school activities in-person.

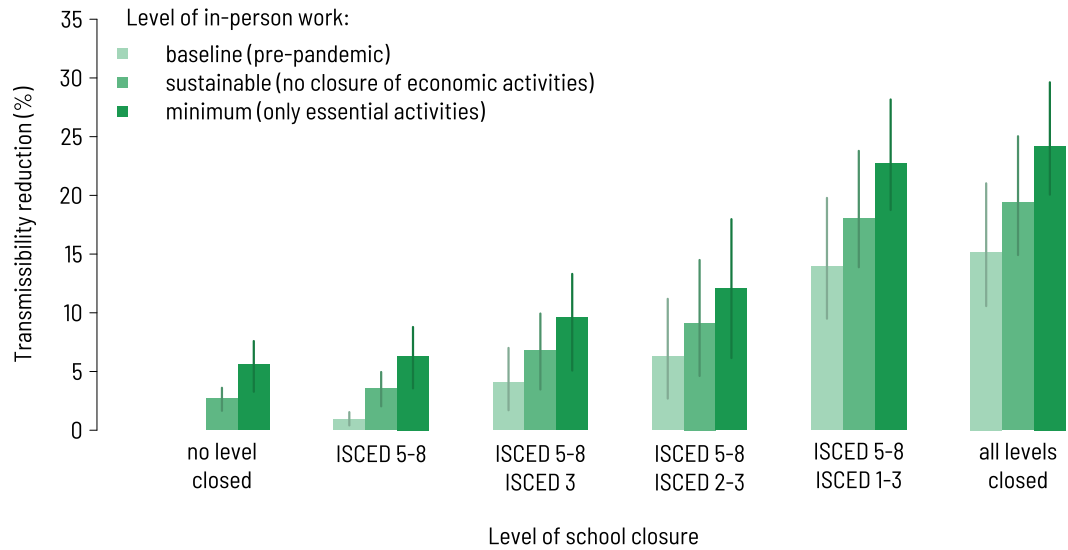

**Figure SI14:** *Effectiveness of in-person school and work closures in reducing viral under scenario 1B assumptions.* The Figure shows the relative reduction in transmissibility associated with intervention scenarios combining different levels of school closures and in-person work, compared to a baseline scenario where the proportion of in-person workers is set at pre-pandemic levels[6] and students across all education levels attend in person.

## 2.5 Sensitivity analysis for social contacts under scenario 2A

### Modelling the determinants of social contacts for adults under scenario 2A assumptions

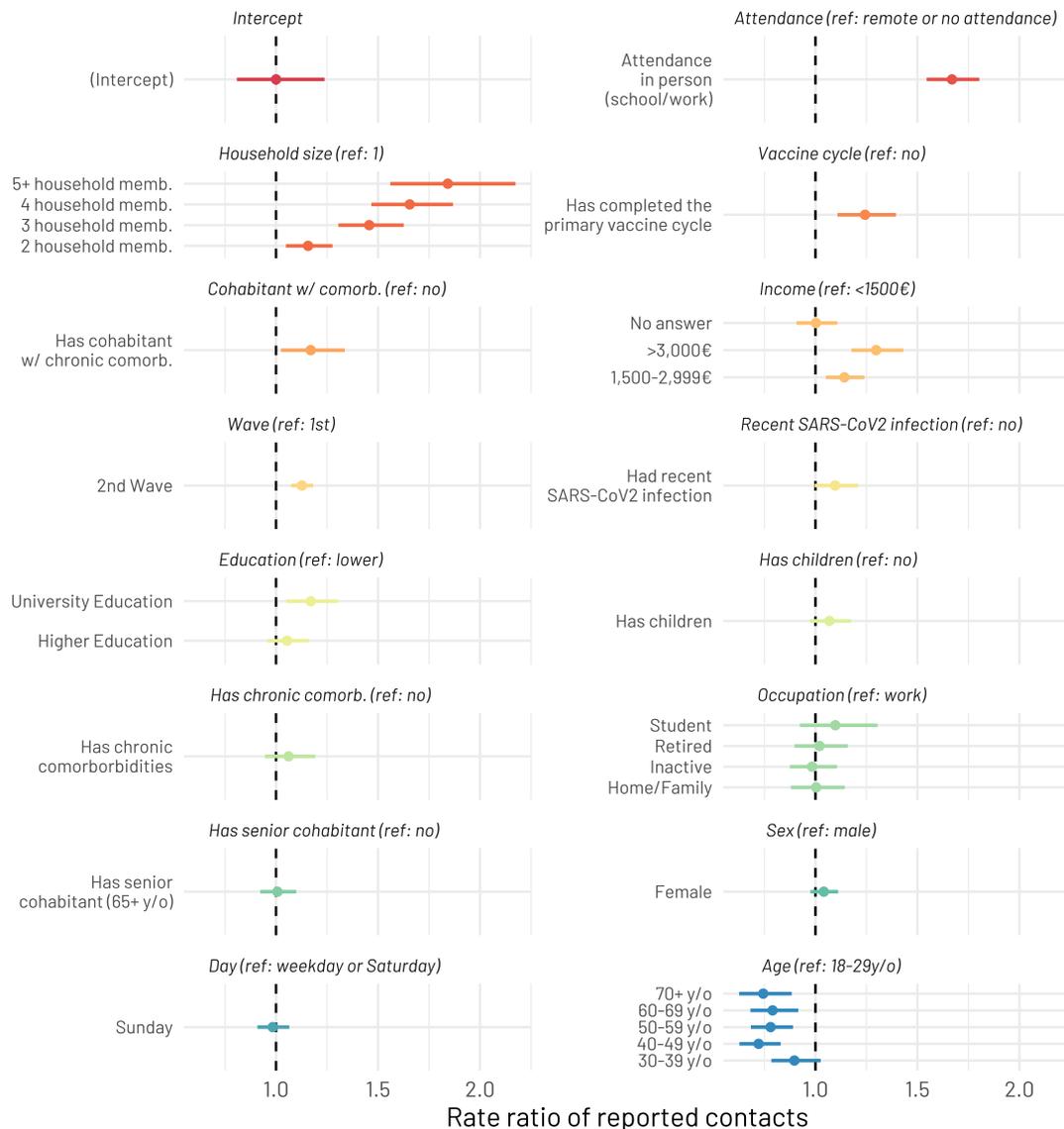

**Figure SI15:** *Model results for the social contacts of the adult population under scenario 2A assumptions.* Each panel in the figure shows the risk ratio of reporting a larger number of social contacts in the adult population for each of the model's covariates.

## Modelling the determinants of social contacts for the underage population under scenario 2A assumptions

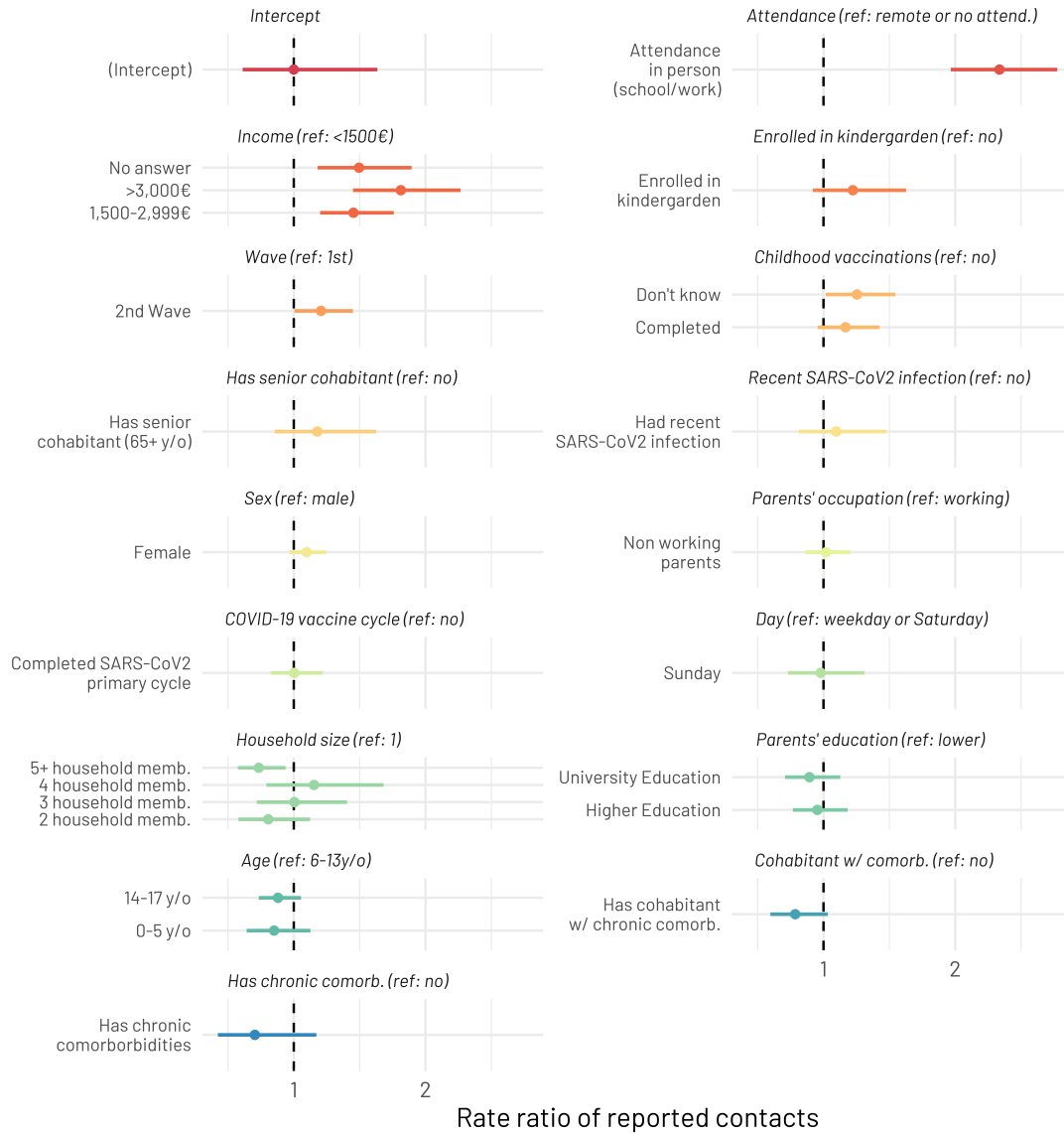

**Figure SI16:** *Model results for the social contacts of the underage population under scenario 2A assumptions.* Each panel in the figure shows the risk ratio of reporting a larger number of social contacts in the underage population for each of the model's covariates.

## 2.6 Additional sensitivity analysis for contact matrices

In the main manuscript analysis, we leveraged the statistical power of both waves to provide a more precise contact pattern description with an age stratification in groups of 5 years. Results from statistical model ensures that the most significant contribution to the number of social contacts is accounted for by correctly weighting the population working or attending school in presence. In this section, we provide additional evidence that the differences between the contact patterns when leveraging data only from the 2023 wave and when pooling both waves together do not show significantly different results in terms of the average number of social contacts by each age-group. Figure SI17 shows for the "overall" population (A), for those working or attending school "in-presence" (B), and for those not attending in-presence (C) the difference between the age-stratified contact patterns from 2023 wave and when pooling both waves together. In the top panels of the Figure we report the average difference in the number of reported social contacts. Shaded areas show the 95% C.I. computed trough 1000 bootstrap iterations. The panels at the bottom shows the element-wise difference between 2023 wave and both waves contact matrices. Note that when only employing 2023 wave data some element of the "not in-presence" matrix have missing values. This is not occurring in the case of poole contact data.

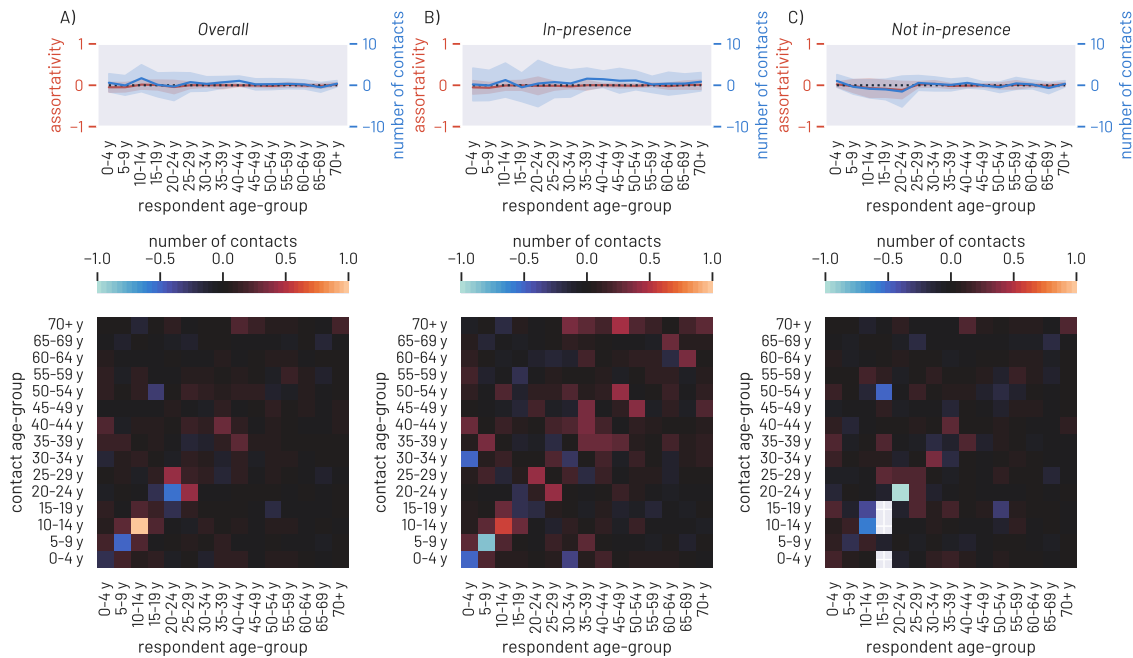

**Figure SI17:** *Difference of 2023 wave contact matrices and both-waves contact matrices for social contacts under scenario 1A assumptions.* A) Overall direct and indirect B1 social contacts, B) Contacts of respondents who attended in-person work or school activities, C) Contacts of respondents who did not attend work or school activities in-person. Gray squares show elements without any contact information available for 2023 wave data.

# 3. Social contact survey: English translation

| Social Contacts Patterns – Italy, March2022 and March2023 |  |
| --- | --- |

| Country | Italy |
| --- | --- |
| **Subject** | Social contacts mixing patterns |

#Demographic info

*Adult Respondent (18+yo)*

| Info | PDL VARIABLE |
| --- | --- |
| Date | |
| gender | gender |
| Age | age |
| Province | GeoPC_Region2_it |
| Education | education_EN |
| Occupation | workstatus_EU |

*Child Respondent (0-17 yo)*

| Info | PDL VARIABLE |
| --- | --- |
| Date | |
| gender | childgenders_EU_1 to childgenders_EU_5 |
| Age | child_ages_mc_2019_EU (computed PDL) |
| Occupation | workstatus_EU =if childage>=16 |

*NOTES:*

- *Whenever questions have "# **Receiver if not adult: parent** ", please insert a warning message in bold-red to let the respondent understand the question is being asked to the parent not the child. Warning message if childage=0to13: " **This question is about you as a parent or guardian of the minor** ", Warning message if childage=14to17: " **This question is about $pipein ["your mom", if gender=2, "your dad" if gender=1], please let $pipein ["she", if gender=2, "he" if gender=1] answer it**"*
- *Whenever respondent_type=1 and "# **Receiver if not adult IS NOT parent** ", we need another warning message (in bold blue color), as follows: "This question concerns ["your child" if childgender=1] ["your daughter" if childgender=2] [$pipe-in:childage] years old"*
- *Added a warning message:*
  *whenever respondent_type =**2** (receiver aged 14-17) (and "# **Receiver if not adult IS NOT parent** ", we need another warning message (in bold blue color), as follows: "This question is about you, ["child" if childgender=1] ["daughter" if childgender=2] of [$pipe-in:childage] years as a recipient of this survey"*
- *Adapt age-specific and gender-specific wording:*
  *Age_gender: wording change: from "Are you…?" to "What is your sex?"*

*[TEST MODE VARIABLE]: based on who we're inviting to answer*
**[respondent_type] respondent**

<1>0-13 yo (via parent)
<2>14-17 yo (kids survey)
<3>Adult (18+)

*# toscripting, INTRO TEXT 1. show only recontacted respondents from wave 1*
*# **respondent_type=3 (adult 18+) version***

*Dear respondent,*

*in March 2022 you were kind enough to respond to a survey conducted by the Bruno Kessler Foundation (Trento) and Bocconi University, on the topic of health.*

*We would greatly appreciate your participation in a second phase of the research and therefore ask you to be available to take part in this new survey. The questions will be the same, precisely because the purpose of the research is to understand if and how people's behaviour evolves over time.*

*Click on ">" to proceed*

---

*#toscripting, INTRO TEXT 1. show only recontacted respondents from wave 1*
*# **respondent_type=1 (0-13yo kid) version***

*Dear respondent,*

*in March 2022 you were kind enough to answer a survey **together to [pipe-in: "your son" if childgender=1 or "your daughter" if childgender=2] (or acting on his behalf), who at the time of completion was [pipe-in number from childage] years old** . The survey , on the topic of health, was conducted by the Bruno Kessler Foundation (Trento) and Bocconi University.*

*We would greatly appreciate your participation in the second phase of the research. We would therefore like **to reinvite** you and **[pipe-in: "your son" if childgender=1 or "your daughter" if childgender=2] a of $pipein** [childage] **years (as of March 2022)** , if possible, to take part in the survey. The questions will be*

*the same, precisely because the purpose of the research is to understand if and how people's behaviour evolves over time.*

We will be asking for $pipe1_kid_gender's opinion on a variety of health-related topics. Many of the questions will be directly about $pipe1_kid_gender, while others will be about you as a parent/guardian. To make it easier to distinguish, [ *red colour*: "questions about you will be marked in red"], while [ *blue colour*: "questions about [pipe-in: "him" if childgender=1, "her" if childgender=2] will be marked in blue"]

***We invite you, if possible, to complete the questionnaire with his/her assistance (where necessary).***

You will be the one to answer, acting as <b>pipe-in: "your son" if childgender=1, "your daughter" if childgender=2] of $pipein *[childage]* years</b>.

*In addition to requesting your consent to proceed, we also need $pipe1_kid_gender to have the will to participate. For this reason, we have created a <u>privacy notice</u> ADD PRIVACY NOTICE that informs you about what we will ask for and how we will use the information.*

*Click on ">" to proceed*

---

*Dear respondent,*

*a few months ago, you and **[pipe-in: "your son" if childgender=1 or "your daughter" if childgender=2] of [pipe-in number from childage] years old at the time of filling out the form** were kind enough to respond to a survey on the topic of health conducted by the Bruno Kessler Foundation (Trento) and Bocconi University.*

*greatly appreciate your participation in the second phase of the research.* We would therefore like to **<u>invite</u>** you and [pipe-in: "your son" if childgender=1, "your daughter" if childgender=2] of $pipein *[childage]* years (in March 2022) to take part in the survey. *The questions will be the same, precisely because the aim of the research is to understand if and how people's behaviour evolves over time.*

<b><u>Many of the questions will be addressed directly to $pipe1_kid_gender </b></u>, ***we encourage you to give** [pipe-in: " **your son** " if childgender=1, " **your daughter** " if childgender=2] **your smartphone/computer/tablet** directly and have [pipe-in: "him" if childgender=1, "her" if childgender=2] fill out the questionnaire directly.*

Some questions will, however, be directed at you as a parent or guardian, so we ask that you remain with $pipe1_kid_gender throughout the entire survey.

To make it easier to distinguish, [ *blue colour*: questions regarding [pipe-in: "your son" if childgender=1, "your daughter" if childgender=2] will be marked in blue], while [ *red colour*: "those regarding you as a parent will be marked in red"].

In addition to requesting your consent to proceed, we also need $pipe1_kid_gender to have the will to participate. For this reason, we have created a <u>privacy notice</u> *[ADD PRIVACY NOTICE image link]* which informs you about what we will ask for and how we will use the information.

*Click on ">" to proceed*

# Questionnaire START (FOR FRESH RESPONDENTS)

**[respondent_type] respondent**
<1>0-13 yo (via parent)
<2>14-17 yo (kids survey)
<3>Adult (18+)

*#Question type: INTERLUDE TEXT FOR **CHILDREN (AGED 0-13)** -> this will be shown if respondent_type = 1*

You recently told us that you have [pipe-in: "a son" if childgender=1, "a daughter" if childgender=2] aged $pipein *[childage]* years living in your household.

We are conducting a survey where we will ask $pipe1_kid_gender's opinion on a range of health-related topics. We would like to invite you and [pipe-in: "your son" if childgender=1, "your daughter" if childgender=2] aged $pipein *[childage]* to take part in the survey. Many of the questions will directly concern $pipe1_kid_gender, while others will concern you as a parent/guardian. To make it easier to distinguish, [ *red colour*: "questions about you will be marked in red"], while [ *blue colour*: "questions about [pipe-in: "him" if childgender=1, "her" if childgender=2] will be marked in blue"]

***We invite you, if possible, to complete the questionnaire with his/her assistance (where necessary).***

You will be the one to answer, acting as <b>pipe-in's representative: "your son" if childgender=1, "your daughter" if childgender=2] aged $pipein *[childage]* years</b>.

In addition to requesting your consent to proceed, we also need $pipe1_kid_gender to have the will to participate. For this reason, we have created a <u>privacy notice</u> *ADD PRIVACY NOTICE image link]* that informs you about what we will ask for and how we will use the information.

*#Question type: INTERLUDE TEXT FOR **CHILDREN (AGED 14-17)** -> this will be shown if respondent_type = 2*

You recently told us that you have a [pipe-in: "a son" if childgender=1, "a daughter" if childgender=2] aged $pipein *[childage]* years living in your household.

We are conducting a survey where we will ask $pipe1_kid_gender's opinion on a variety of health-related topics.

We would like to invite you and [pipe-in: "your son" if childgender=1, "your daughter" if childgender=2] aged $pipein *[childage]* to take part in the survey. <b><u>Many of the questions will be directed to $pipe1_kid_gender </b></u>, ***we encourage you to hand your smartphone/computer/tablet directly to*** [pipe-in: " ***your son*** " if childgender=1, " ***your daughter*** " if childgender=2] and have [pipe-in: "him" if childgender=1, "her" if childgender=2] fill out the questionnaire.

Some questions will, however, be directed at you as a parent or guardian, so we ask that you remain with $pipe1_kid_gender throughout the entire survey.

To make it easier to distinguish, [ *blue colour*: questions regarding [pipe-in: "your son" if childgender=1, "your daughter" if childgender=2] will be marked in blue], while [ *red colour*: "those regarding you as a parent will be marked in red"].

In addition to requesting your consent to proceed, we also need $pipe1_kid_gender to have the will to participate. For this reason, we have created a <u>privacy notice</u> *[ADD PRIVACY NOTICE image link]* which informs you about what we will ask for and how we will use the information.

*#Base: Children aged 14-17 yo*
*#Filter: childage =14to17*
*#Receiver if not adult: parent*
*#Question type: single*

**[child1417_presence] Are you available and willing on the part of [pipe-in: "your son" if childgender=1, "your daughter" if childgender=2] aged $pipein *[childage]* to participate in this survey?**
***Please select one response only* .**

<1>Yes, [pipe-in: "my son" if childgender=1, "my daughter" if childgender=2] is available to answer the survey now.

<2>Yes, but [pipe-in: "my son" if childgender=1, "my daughter" if childgender=2] CANNOT answer the survey now

<3>No, I don't want [pipe-in: "my son" if childgender=1, "my daughter" if childgender=2] to answer the survey.

*#TOSCRIPTING: screenout if child1417_presence=3, go to page "notthere" if child1417_presence=2*

*#Base: Children aged 14-17 yo*
*#Filter: childage =14to17 and child1417_presence=2*
*#Receiver if not adult: parent*
*#Question type: interlude page*

**[notthere] You said that [pipe-in: "your son" if childgender=1, "your daughter" if childgender=2] from $pipein *[childage]* is not currently present (or unavailable to respond). You can resume the survey when [pipe-in: "your son" if childgender=1, "your daughter" if childgender=2] is available by clicking the invitation link you received by email, which will take you back to this page.**
**Remember to read the privacy notice and when [pipe-in: "your son" if childgender=1, "your daughter" if childgender=2] is available to respond, click the " > " arrow at the bottom.**
***Select only one answer* .**

" > " continue button

*#Question type: INTERLUDE TEXT **FOR ALL***

Dear respondent,

*#skip this paragraph if respondent=recontact*
in the context of a research project carried out by the Bruno Kessler Foundation (Trento) and Bocconi University (Milan) we are conducting a survey on the topic of health.
The aim of the research is to deepen our knowledge of the interactions between people that carry the transmission of infectious diseases.

<br/>You will be credited $points points for completing the survey.<br/>We have tested the questionnaire and it takes on average about $time_var minutes to complete.

If necessary, you can interrupt the questionnaire and resume later.<p></p><b>The questionnaire is completely anonymous and you will NOT be identified IN ANY WAY</b> [#t **oscripting** : rewording of this sentence from "the questionnaire" to "identified" for **respondent_type=1** : "The questionnaire is completely anonymous and neither you nor $pipe1_kid_gender will be identified IN ANY WAY"]. Please complete the entire questionnaire, including the questions that help us better understand your profile. The success of the research depends on the completeness of the answers given by each individual respondent.<p></p>Your contribution is greatly appreciated! We thank you very much for participating in our survey.

*#TOSCRIPTING: show if respondent_type=3*
<i>We will take all precautions to ensure that your anonymity is preserved during every phase of the research. Further information is available on the <a href="…" target="_blank">privacy and cookies notice</a> page. <br/><br/>

By clicking ">" you confirm that you wish to participate in the survey. </i><br/><br/>

*#Base: 0-5 yo*
*#Filter: if childage =0to5*
*#Receiverif not adult : parent*
*#Question type: single*
*#alternativetextif childage =5 "[pipe-in: "Your child" if childgender =1, "Your daughter" if childgender =2] of $ pipein [ childage ] years goes to kindergarten or school? "*

**[occupation_05] [pipe-in: "Your son" if childgender=1, "Your daughter" if childgender=2] of $pipein**
***[childage] years old goes to kindergarten (daycare or preschool)?***

<1>Yes
<2>No

*#Base: All*
*#Filter:*
*#Receiver if not adult : child if respondent_type = 2 (and childage = 16,17), parent if respondent_type =1*
*or if respondent_type =2 (and childage =14,15). Adult him/ herslef if receiver_type =3.,*
*if respondent_type_LF =2 -> In case of respondent_type =2 (and childhood =14,15), " Warning*
*message if childage=14to17" needs to appear as per word page 1 instructions*
*if respondent_type =2 and childage =16,17 insert blue bold warning message "This question concerns*
*you, ["child" if childgender=1] ["daughter" if childgender=2] of [$pipe-in:childage] years old as the*
*recipient of this survey"*
*if respondent_type =2 and childage =16,17 AND code selected IS NOT 5: display warning message:*
*"You are under 18 years old and have stated that he or she does not attend school . This answer is*
*plausible, but please review it to ensure it is accurate. Please check your answer and continue."*
*#Question type: single*
*#code 4 amend italian wording ( without ending point): "I am a househusband, I take care of the family"*
*#if RESPONDENT=RECONTACT pre-fill the answer option with wave 1 code selection and pipe-in text*
*below all question text ( bold and underscored): "A few months ago you selected the answer below, if*
*it is still valid, continue without changing your answer, otherwise change it by selecting the relevant*
*option"*

**#TOSCRIPTING: INSERT PDL " workstatus_EU " and rename it [occupation]**
**Type in the question wording of pdl , please amend : "Which of these options best describes your**
**employment situation?"**

**[ employment ]** {single varlabel ="work status"} Which of these BEST applies to you

<1>Working full time (30 or more hours per week) including temporarily off work
<2>Working part-time (8 to 29 hours per week) including temporarily off work
<3>Working part-time (less than 8 hours a week including temporarily off work)
<4>Looking after the home or family
<5>Full-time education
<6>Part time Student (working more than 30 hours per week)
<7>Part time Student (working less than 30 hours per week)
<8>Retired from paid work
<9>Unemployed

<10>Not working due to long term illness/incapacity/disability
<11>Not working for other reasons
<96> Other

---

*#Base: Employed 16+ yo*
*#Filter: if occupation = (or)1,2,3,6,7 (deleted code 4) and age or childage >=16*
*#Receiver if not adult: child (if childage >=16), parent if childage <=15, if respondent_type_LF =2*
*if respondent_type =2 and childage =16,17 insert blue bold warning message "This question is about you, ["child " if childgender=1] ["daughter"if childgender=2] of [$pipe-in:childage] years old as the recipient of this survey"*
*#Question type : single, ~~randomize order~~*
*#if RESPONDENT=RECONTACT pre-fill the answer option with wave 1 code selection and pipe-in text below all question text ( bold and underscored ): "A few months ago you selected the answer below, if it is still valid, continue without changing your answer, otherwise change it by selecting the relevant option"*

**[ occupation_category | Which job category best describes your primary occupation?**
***Select one answer.***
<1>Agriculture, livestock, fishing and forestry
<2>    Hotel and catering
<3>    Administrative and service support activities
<4>    Arts, Entertainment and Recreation
<5>    Activities of non-territorial bodies and organisations
<6>    Domestic work activities
<7>    Real estate activities
<8>    Scientific-technical professional activities
<9>    Wholesale and retail trade, repair of motor vehicles and motorcycles
<10>   Construction
<11>   Mining and quarrying, mining industry
<12>   Finance and insurance
<13>   Computer science and communication
<14>   Instruction
<15>   Logistics, transport and storage
<16>   Manufacturing
<17>   Public Administration, Defense, Compulsory Social Security
<18>   Electricity, gas, steam and air conditioning resources
<19>   Water resources, sewerage, waste management and remediation activities
<20>   Health and social services
<21>   Other services
<955 fixed > Other occupation not listed above *[SPECIFY]*

---

*#Base: Total*
*#Filter:*

**[instruction]**
#TOCLIENT: we insert our database variable that will be recoded as per request into [middle school diploma or less, high school diploma, degree or more]
#TOSCRIPTING: education =1 'lower' if education_level_IT =1,2,3; education =2 'mid' if education_level_IT =4,5,6, education =3 'higher' if education_level_IT =7to12, education =4 'other/no answer' if education_level_IT =13,14

[ education_level_it ] {single} What is the highest level of education or qualification you have achieved?
<1> No diploma
<2> Elementary school diploma
<3> Middle school diploma
<4> High School Diploma
<5> Apprenticeship Diploma
<6> Technical Institute
<7> Bachelor's Degree
<8> Single-cycle master's degree
<9> Old system degree
<10> First level Master
<11> Specialist degree/ Master's degree
<12> Doctorate / second level master's degree
<13> Other
<14> I prefer not to answer

**[income] FAMILY income is the combined income of all earners in a household from all sources (including wages and annuities). What is the NET monthly household income (i.e. excluding taxes) of your household?**

***If you live with people who are not part of your household, exclude them from the calculation.***

***Select one answer only.***
<1> < EUR 1,000
<2>     EUR 1,000-1,499
<3>     EUR 1,500-1,999
<4>     EUR 2,000-2,499

| | |
|---|---|
| <5> | EUR 2,500-2,999 |
| <6> | EUR 3,000-3,499 |
| <7> | EUR 3,500-3,999 |
| <8> | EUR 4,000-4,999 |
| <9> | > EUR 5,000 |
| <933> | I prefer not to answer |

*#Base: Total*
*#Filter:*
*#Receiver if not adult: if childage between 0 and 13: parent - if childage between 14 and 17: child*
   *if respondent_type = 2 insert blue bold warning message "This question is about you, ["son" if*
   *childgender=1] ["daughter"if childgender=2] aged [$pipe-in:childage] as the recipient of this*
   *survey"*
*#Question type : Single*

**[ perceived_income ] From 1 to 5, what level of economic well-being would you place yourself in**
   **compared to the Italian population?**
   ***Select only one answer.***

| | |
|---|---|
| <10> | 1 = very low well-being |
| <11> | 2 = low well-being |
| <12> | 3 = average well-being |
| <13> | 4 = high well-being |
| <14> | 5 = very high well-being |
| <933> | I prefer not to answer |

**#to scripting: BEFORE ASKING ANY QUESTION ON COHABITANTS (ie before**
   **profile_household_size_EU ), we need a complex pipe in exercise in order to remind the**
   **respondents their cohabitants (as stated by themselves in wave 1)**

*Please show a page with the below text (it means "you declared in March 2022 you live with…"*

*#Base: Total*
*#Filter:*
*#Receiver if not adult: child if respondent_type =2, adult if respondent_type =1*
*#Question type : MULTIPLE*

**[coviventi_check]The people listed below are those you declared you were living with in March 2022.**
   **Please read carefully and answer the following question:**

*#[show a list where you pipe-in age (coviventi_eta) and c_coviventi_name]*
*#list example below:*
*John of 8 yo*
*Wife of 45 yo*
*….*

**Do you still live with all of the above people?** ***Select the appropriate option(s).***

<1>Yes, there have been no changes ~~or new people added~~ *[EXCLUSIVE] [IF SELECTED SKIP ALL THE COHABITANS QUESTIONS]*

<2>One or more of the above persons no longer live with me

<3>I live with one or more new people *[if selected GO TO….]*

<4>None of the above people are my cohabitants anymore *[EXCLUSIVE] [IF SELECTED ASK ALL THE COHABITANS QUESTIONS]*

---

*#Base: Who has at least one less cohabitant*
*#Filter: if [ cohabitants_check ]= 2*
*#Receiver if not adult: child if respondent_type =2, adult if respondent_type =1*
*#Question type : multiple - required soft*

**[cohabitants_check_2] Which of the following people <b>NOT</b> live with you anymore? *Select all that apply.***
***If you made a mistake in the previous question and all the people below still live with you, proceed without selecting any people.***

*#[show a list where you pipe-in age (conviventi_eta) and c_covienti_name] and make multiple answer options selectable*
*# →cohabitants that were selected in conviventi_check_2 will have to be removed from c_conviventi (ie they should not be considered cohabitants anymore)*

---

*#Base: Who has at least one more cohabitant*
*#Filter: if [ cohabitants_check ]= 3*
*#Receiver if not adult: child if respondent_type =2, adult if respondent_type =1*
*#Question type : single - required soft*

**[conviventi_check_3] How many new people do you live with in addition to those listed below? *Select only one answer option.***
***If you made a mistake in the previous question and do not live with any new people, proceed without selecting an answer.***

*#[show a list where you pipe-in age (conviventi_eta) and c_covienti_name] – NOT SELECTABLE*

<1>1
<2>2
<3>3
<4>4
<5>5

*# →if code1,2,3,4,5 is selected re-ask conviventi_eta and conviventi_name just for the new cohabitants (1,2,3,4 or 5 depdenting on code selected in conviventi_check_3)*
*# →after the respondent will have indicated these new cohabitans age (conviventi_eta) and name (conviventi_name) they have to be added to c_conviventi list (ie added to the new cohabitants list)*

**COMPUTE A VARIABLE [CONVIVENTI_excl_parentANDunderage] THAT IS =**
**[PARTNERS] number (excl code 98 and 95) – [profile_household_child_respon_EU] number (code**
**number -1, up to code 6 only) – 1 (parent him/herself)**

*#Base: 0-13yo*
*#Filter: if cohabitants >1 AND respondent_type =1*
*#Receiver if not adult: PARENT (SHOW WARNING MESSAGE)*
*#Question type: single grid*
*#TOSCRIPTING: valid for '' conviventi_eta_013yo to conviventi_eta_18+'', client noticed that a pipe-in is*
*in place not only for rows _0 and _1 (or just for row _0 in conviventi_eta_18+), but also for other*
*possible children (take from childhoods pdl) . That could be ok, and facilitate respondents answer, BUT*
*as of now we have some ''bug'': eg A parent has two underage children and lives in a household of 5*
*people including him/herself: now have two pipes-in (for the underage children) and three empty*
*spaces. We should now have 3 pipes-in (1 is the parent, 2 are for the underage children)*
**[conviventi_eta_013yo] How old are the people you live with?**
    ***Select an answer for convivente.***

    ***pipes-in info (not counted in rows)***

    You *pipe-in VISIBLE to respondent: this is age from parent panellist* years
    "Your son involved in this survey/Your daughter involved in this survey " *(according to [childgender]=1 or 2*
    *of **receiver** ) + pipe in from [childage] – visible to respondent if cohabitants>1* years
    "Your son/daughter " *[pipe in [childage]], as many pipe-in here as we have in*
    *profile_household_child_respon_EU (-1 of the child receiver)*

    **-[cohabitants_age_013yo_1** ] Other cohabitant *as many rows (_2, _3 and so on) a the number of this*
    *computed variable [CONVIVENTI_excl_parentANDunderage]*
    [open integer] *(min 0 max 110)*

*#Base: 14-17yo*
*#Filter: if cohabitants >1 AND respondent_type =2*
*#Receiver if not adult: child*
    *if respondent_type = 2 insert blue bold warning message "This question is about you, ["son" if*
    *childgender=1] ["daughter"if childgender=2] aged [$pipe-in:childage] as the recipient of this*
    *survey"*
*#Question type : single grid*
**[conviventi_eta_1417yo] How old are the people you live with?**
    ***Select an answer for cohabitant.***

***pipes-in info (not counted in rows)***

    You *pipe-in VISIBLE to respondent: this is age from parent panellist* years
    "Your father involved in this survey/Your mother involved in this survey " *(according to [childgender]=1 or*
    *2 of **receiver** ) + pipe in from [childage] – visible to respondent if cohabitants>1* years

"Your brother/sister/partner " *[pipe-in [childage]], as many pipe-in here as we have in profile_household_child_respon_EU (-1 of the child receiver)*

-**[cohabitants_age_1417yo_1]** Other cohabitant *as many rows (_2, _3 and so on) a the number of this computed variable [CONVIVENTI_excl_parentANDunderage]*

[open integer] *(min 0 max 110)*

*#Base: adults 18+*
*#Filter: if cohabitants >1 AND respondent_type =3*
*#Receiver if not adult:*
*#Question type: single grid*
*#TOSCRIPTING: pipes-in for underage children ( childage pdl)*

Patterns- Fitting Predicted Matrices to Serological Data. *PLOS Comput. Biol.* **6**, e1001021 (2010).

*apply here too*

**[cohabitants_age_18+] How old are the people you live with?**
    ***Select an answer for cohabitant.***

*pipes-in info (not counted in rows)*
  You *pipe-in VISIBLE to respondent: this is age from parent panelist* years
  "Your son/daughter " *[pipe-in [childage]], as many pipe-in here as we have in*
  *profile_household_child_respon_EU*

  **-[cohabitants_age_18+_1]** Other cohabitant *as many rows (_2, _3 and so on) a the number of this computed*
  *variable [CONVIVENTI_excl_parentANDunderage]*
  [open integer] *(min 0 max 110)*

---

*#Base: Total*
*#Filter: if aft least one "Other cohabitant" is shown in conviventi_eta_xxx question*
*#Receiverif not adult :* PARENT if respondent_type =1, child if respondent_type =2
  *if respondent_type = 2 insert blue bold warning message "This question concerns you, ["son" if*
  *childgender=1] ["daughter"if childgender=2] aged [$pipe-in:childage] as the recipient of this*
  *survey"*
*#Question type: open text boxes*

**[c_coviventi_name] Enter the name of each <u>cohabitant</u> listed below. <u>Only one name per box</u>.**
    ***You do not need to enter their real name; it can also be a nickname that will help you distinguish***
    ***these people in the next phase of the questionnaire.***

**[in rows]:** *as many rows as "Other cohabitant" as we have in conviventi_eta_xxx question*
    *[pipe-in: name wording from conviventi_eta_xxx] + "di" + "pipes-in: age from conviventi_eta" +*
    *"anni"*
**[In columns]:** *open text boxes*

*#toscripting: c_coviventi_name (as written in open boxes here) will have to appear in <u>closed to</u>*
    <u>c_vaccino</u> *questions as it happens for names that are filled in c_home to c_otheroutdoor questions:*
    <u>Note</u> : *please update on timing*

*#to scripting, changed 4 questions order:* **THESE are TO BE KEPT AT THE END after income received**
- *implicit_acute_pathologies*
- *implicit_chronic_pathologies*
- *implicit_pathologies_acute_cohabitant*
- *implicit_patologies_croniche_cohabitant*

*#Base: Total*
*#Filter:*
*#Receiverif not adult : child - if respondent_type_LF =1 ->*
   *if respondent_type = 2 insert blue bold warning message "This question concerns you, ["son" if childgender=1] ["daughter"if childgender=2] aged [$pipe-in:childage] as the recipient of this survey"*
*#Question type: multiple, randomize rows – skip option health*
**[ acute_pathologies ] Are you currently experiencing one or more of the following symptoms?**
   ***Select all that apply.***

<1>No clinical condition *fixed and exclusive (please note: delete former code 2: no clinical condition, that was reapeted from code 1)*

<2>Chills

<3>General weakness

<4>Diarrhea

<5>Difficulty breathing

<6>Fever (37.5°C or higher)

<7>Sore throat

<8>Muscle ache

<9>Loss of taste

<10>   Loss of smell

<11>   Pneumonia

<12>   Confusional state or delirium

<13>   Cough

<14>   Vomit

<955 *fixed* >Other clinical conditions

*#Base: Total*
*#Filter:*
*#Receiverif not adult : child , if respondent_type_LF =1 ->*
   *if respondent_type = 2 insert blue bold warning message "This question concerns you, ["son" if childgender=1] ["daughter"if childgender=2] aged [$pipe-in:childage] as the recipient of this survey"*
*#Question type: multiple, randomize rows – skip option health*
**[ chronic_pathologies ] Indicate if you have one or more of the following pathologies or are in one of the following conditions?**
   ***Select all that apply.***

<1>No pathology *fixed and exclusive*

<2>   Alcohol or other drug abuse *shown if respondent_type = 2,3*

<3> Other immunodeficiencies (e.g. transplant patient, long-term use of immunosuppressant drugs)
*fixed above code 955*
<4> Sickle cell anemia or thalassemia
<5> Heart conditions (e.g., heart failure, coronary artery disease, cardiomyopathy, or hypertension)
<6> Neurological conditions
<7> Diabetes (type 1 or type 2)
<8> Smoker *shown if respondent_type = 2.3*
<9> Pregnancy *fixed above code 955 – show if : gender=2(female) and if respondent_type = 3*
<10> Stroke or cerebrovascular disease
<11> HIV/AIDS infection
<12> Chronic liver disease
<13> Chronic kidney disease
<14> Chronic lung diseases (e.g. COPD, asthma, cystic fibrosis, etc.)
<15> Overweight and obesity
<16> Tumors
<955 *fixed* >Other clinical conditions

*#Base: Who lives with someone*
*#Filter: if cohabitants>1*
*#Receiver if not adult: parent - if respondent_type_LF =1 -> child if respondent_type_LF =2 (rewording not needed)*
   *no blue warning here for resp type=1,2*
*#Question type: multiple, randomize rows – skip option health*
   *#toscripting: we miss underline in cohabitant*
**[ pathologies_acute_cohabitant ] Indicate whether <u>at least one of your cohabitants</u> has one or more of the following symptoms these days.**
***Select all that apply.***
<1>No clinical condition *fixed and exclusive*
<2>Chills
<3>General weakness
<4>Diarrhea
<5>Difficulty breathing
<6>Fever (37.5°C or higher)
<7>Sore throat
<8>Muscle ache
<9>Loss of taste
<10> Loss of smell
<11> Pneumonia
<12> Confusional state or delirium
<13> Cough
<14> Vomit
<955 *fixed* >Other clinical conditions
<977 *fixed* >I don't know

**[ cohabitant_chronic_pathologies | Indicate whether <u>at least one of your cohabitants</u> has one or more
of the following pathologies or is in one of the following conditions.**
***Select all that apply.***

<1>No pathology *fixed and exclusive*

<2>Abuse of alcohol or other drugs

<3>Other immunodeficiencies (e.g. transplant patient, long-term use of immunosuppressant drugs) *fixed
    above code 955*

<4>Sickle cell anemia or thalassemia

<5>Heart conditions (e.g. heart failure, coronary artery disease, cardiomyopathy, or hypertension)

<6>Neurological conditions

<7>Diabetes (type 1 or type 2)

<8>Smoker

<9>Pregnancy *fixed above code 955*

<10>   Stroke or cerebrovascular disease

<11>   HIV/AIDS infection

<12>   Chronic liver disease

<13>   Chronic kidney disease

<14>   Chronic lung diseases (e.g. COPD, asthma, cystic fibrosis, etc.)

<15>   Overweight and obesity

<16>   Tumor

<955 *fixed* >Other pathologies

<977 *fixed* >I don't know

**[covid] Have you ever contracted COVID-19, diagnosed by antigen or molecular swab?**
***Select only one answer.***

<1>Yes

<2>No, never

**[ covid_date ] When were you diagnosed with COVID-19?**
***If you have had it multiple times, think about the last time. Select only one answer by clicking the box below.***

[FIRST DROP DOWN: **MONTHS+YEAR** *FROM: March 2020 to SAME DAY OF INTERVIEW* ]
[SECOND DROP DOWN: **DAY** , *from 0 to 31 according to months (exclude numbers on relevant months) + Don't Know option* ]

**[ covid_ospedale ] Have you \<b\> EVER \</b\> been hospitalized for COVID-19?**
***Select one answer only.***

<1>Yes
<2>No, never

**[info] Which of the following media do you regularly use as a source of information?**
***Select all that apply.***

<1>Paper newspapers or magazines
<2>    Online newspapers or magazines (including apps and blogs)
<3>    Television
<4>    Radio (live)
<5>    Podcast or radio podcast
<6>    Instagram
<7>    Facebook
<8>    Twitter
<9>    Whatsapp (for example through "groups")
<10>    Telegram
<11>    Tik Tok
<12>    YouTube
<955 fixed>Other (not listed above) *[SPECIFY]*

## #Module 2A: Questions about "YESTERDAY"

*#Interludetext: txt4*

The next questions refer to events that happened YESTERDAY.

**We kindly ask you to complete the questions in this section today** .

*#TOSCRIPTING: made the above part bold – wording updated*

*#Base: Total*
*#Filter*
*#Receiverif not adult : child , if respondent_type_LF =1 ->*
*if respondent_type = 2 insert blue bold warning message "This question concerns you, ["son" if childgender=1] ["daughter"if childgender=2] aged [$pipe-in:childage] as the recipient of this survey"*
*#Question type: single – skip option health*

**[isolation] Yesterday, were you in…?**
***Select one answer only.***

<1>Fiduciary isolation (for positivity)

<2>    Self-surveillance (as per the legislative decree of 30 March 2022)

<3>    At home due to illness (but not in isolation due to positivity)

<4>    None of these options

*#Base: Employed 16+ yo not in isolation/quarantine or on sick leave YESTERDAY*
*#Filter: if ( employment = (or)1,2,3, ,6,7 (deleted code 4) and age or childage >=16) AND isolation != 1,3*
*#Receiver if not adult: child (if childage >=16)*
*if respondent_type = 2 insert blue bold warning message "This question concerns you, ["son" if childgender=1] ["daughter"if childgender=2] aged [$pipe-in:childage] as the recipient of this survey"*
*#Question type: single, randomize ($RAND=2)*

**[ presence_work ] What mode did you work in YESTERDAY?**
***Select one answer only.***

<1>I have worked exclusively remotely

<2>    I have worked entirely or partly in person

<3>    I didn't work yesterday *#TOSCRIPTING: FIXED*

*#Base: Students not in isolation/quarantine or on sick leave YESTERDAY*
*#Filter: if [(age/ childage >=16 AND occupation = (or)5,6,7) OR childage =6to15 OR ( childage =0to5 AND occupation_05=1)] AND isolation !=1,3*

*date (new dates added):in case respondent type = 2 (14-17 yo )*
*date is 15.march to 20.march + 22.march to 27.march + 29.march to 03.april + 05.april to 10.april (Tuesdays to Sundays)*

*in case respondent type =1,3 (0-13yo or 18+) date is 15.march to 19.march + 22.march to 26.march + 29.03 to 02.april + 05.april to 09.april (Tuesdays to Saturdays)*

*#Receiver if not adult: child, if respondent_type_LF =1 ->*
*if respondent_type = 2 insert blue bold warning message "This question concerns you, ["son" if childgender=1] ["daughter"if childgender=2] aged [$pipe-in:childage] as the recipient of this survey"*
*#Question type: single randomize ($RAND=2)*

**[ presence_school ] What type of teaching did you do YESTERDAY?**
***Select one answer only.***

<1>Exclusively remotely (distance learning/DAD)

<2>     In whole or in part in presence

<3>     None (e.g. school/university was closed, I was sick) *#TOSCRIPTING: FIXED*

*#TOSCRIPTING: question order: please leave c  home first, then randomize c  school and c_work as second, randomize other questions in third place, finally randomize c_other_indoor and c_other_outdoor in last place.*

c_otherindoor, c_otheroutdoor, c_restaurant, transport, leisure: max 10 each
c_shopping, c_homeguest, c_home, c_work, c_school: max 20 each

*#TOSCRIPTING: In any circumstances, for the questions **c_home to c_otheroutdoor** we need to NOT allow respondents to use a same exact name within the same questions and also between these questions. Eg we can never have a duplicate name like "John husband" twice.*
*I confirm this is required even if it's only applicable to same exact wording like "john" and "john"warning message in Italy n: "You have already written the name [pipe-in: duplicated name(s)] once. Please review your answer."*

*#Question type: INTERLUDE TEXT if respondent_type_LF =1 ->*

**Now we will ask you some questions about the <u>close contacts</u> you had <u>YESTERDAY</u> .**

**Please read the following carefully.**

A **CLOSE CONTACT** is defined as:

a) a **physical interaction** with another person (e.g. a handshake, a pat on the back, a hug, a kiss, etc. ); or

b) an interaction with a **person physically present** in which at least **5 words are exchanged** (for example: "Hello!" "Hello!" "How are you?" "Good and you?"). In this case it is not necessary that there has been physical contact with the person.

#TOSCRIPTING: added bold in green parts
please create a drop-down box visible in each question from c_home to c_otheroutdoor question, to let the respondent take a second look, if they feel they need to, at the definition of " contact "CLOSE CONTACT " from this interlude
(ie "A CLOSE CONTACT is defined as:

a) a physical interaction you had with another person (e.g. a handshake, a pat on the back, a hug, a kiss, etc. ); or

b) an interaction with a physically present person in which you exchanged at least 5 words (for example: "Hello!" "Hi!" "How are you?" "Good and you?"). In this case it is not necessary that there was physical contact with the person.")

#TOSCRIPTING: as done for the pop up message "Click here to review the definition of "close contact".
Client also asked to have another pop up… if two pop ups are too much and cause crashes, we can condense working into one single pop up. The wording link for pop up would change into: "Click here to review the definition of "close contact" and reread the instructions on how to enter names"

**[c_conviventi] Indicate which of the following people you live with have had <b>close contact</b> <u>at home</u> YESTERDAY and which have not.**
  *#toscripting: resp_type=1 rewording: see excel file*

*#toscripting:*
  **[ In rows]:** *in case of You, Your son/daughter: same "pipes-in" as in conviventi_eta_xxx questions (names + age numbers pipes-in). In case of "other cohabitant" in conviventi_eta_xxx, please pipe-in written name from [c_conviventi_name] and age from conviventi_eta_xxx . Please exclude the receiver pipe-in: receiver will be excluded from pipe-in rows: if respondent_type=2,3 they will not see "You", if respondent_type=1 (0-13yo) they will not see "My son/daughter involved in this survey"). Logic is we weed all cohabitants to appear, excluding the receiver of the survey.*

**Example for respondent_type=1 rows**

<1><u>I had close contact yesterday (</u>physical contact or exchange of at least 5 words)
  *rewording: see excel file*
<2><u>I didn't have </u>close contact yesterday
  *rewording: see excel file*

**In the following questions we will ask you to list all the <b>people with whom you had close contact YESTERDAY in different places</b> (for example: at your home, at other people's homes, at school/work, in shops, on public transport, etc.).**

Identify each <b>person – separately</b> - with a <u>generic name (<i>for example: friend, colleague, plumber, shop assistant</i>)</u>, <u>you will need to distinguish them in the next phase of the questionnaire</u> .

<b>ATTENTION</b>, the generic name must be unique ( <u>you cannot repeat the same name twice</u> ), if necessary you can differentiate people by adding a name or initials (<i>for example: friend Francesco, colleague Anna, etc.</i>).

*#toscripting: if respondent is child (childage=5to13)*
NOTE: *If [pipe-in: "your son" if childgender=1, "your daughter" if childgender=2] is not* [pipe-in: "stato" if childgender=1, "stata" if childgender=2] *present so far, we invite you to answer the next questions with its support.*

---

*#Base: Total*
*#Filter*
*#Receiver if not adult: child, if respondent_type_LF =1 ->*
*    NO WARNING MESSAGE (neither for resp type =1 nor 2)*
*#Question type : single*
[contact_compilation] This is a TEST question that helps us to fill out the questionnaire CORRECTLY.

In the next questions we will ask you to list the people with whom you have had close contact in different contexts. For example, if you have had close contact with 4 people (Mario, Sara, Mario's son and Sara's daughter), you will have to fill in 4 boxes , as shown in the image below.
*#toscripting: place image in this position*

Now answer this question: <u>How many people you have had close contact with can you fit in each box</u> ? *Select one answer only*

<1>Only one person per box
<2>Up to two people per box
<3>How many people do I want per box?

*#toscripting: **warning message** if code is not 1: "In the next questions, in which you will have to list the people with whom you had close contact yesterday, YOU WILL HAVE TO INSERT ONE PERSON PER BOX", correct your answer to continue". **Do not let respondent proceed if** keep selecting code 2 or 3.*

*#Base: Total*
*#Filter*
*#Receiverif not adult : child , if respondent_type_LF =1 ->*
*if respondent_type = 2 insert blue bold warning message "This question concerns you, ["son" if childgender=1] ["daughter"if childgender=2] aged [$pipe-in:childage] as the recipient of this survey"*
*#Question type: open text boxes,*
**[ c_home ] Think of all the <b> people who DO NOT LIVE with you </b> with whom you have had <b> close contact YESTERDAY AT YOUR HOUSE </b> and list them below <b> one per box </b> .**
***(for example friends or relatives who came to visit, neighbors, boiler man, postman, plumber, cleaner, etc.).***
*[rewording to update]*

[possible contact 1]

[possible contact 2]

[possible contact 3]

[possible contact 4]

[possible contact 5]

    … etc …

<966> *Not applicable: I did not have any close contact at my home with non-cohabiting people yesterday MOVE ABOVE*

*#Base: Who work and have not worked remotely yesterday*
*#Filter: if presence_work =2*
*#Receiverif not adult : child ( only when childage >=16)*
*if respondent_type = 2 insert blue bold warning message "This question concerns you, ["son" if childgender=1] ["daughter"if childgender=2] aged [$pipe-in:childage] as the recipient of this survey"*
*#Question type: open text boxes*
**[ c_work ] Think of all the <b> people </b> you had <b> close contact with YESTERDAY AT WORK </b> and list them, <b> one per box </b>, in the boxes below.**
***If you had close contact with a significant number of people during your work day (e.g. customers for a clerk/front-office worker or students for a teacher), list only the most relevant ones (interactions lasting more than 15 continuous minutes).***
***Note: if there are any of these people that you have already included previously, DO NOT include them in the list below . #Toscripting: this part "exclude contact from the list below" has to be removed: in all these c_xxx questions***

[possible contact 1]

[possible contact 2]

[possible contact 3]

[possible contact 4]

[possible contact 5]

    … etc …

<966> *No new contacts to report MOVE ABOVE*

*#Base: All students who went to school yesterday*

**[ c_school ] Think about all the </b> people </b> you have had <b> <u>close contact</u> with <u>YESTERDAY</u> AT <u>SCHOOL</u> </b> (** *including nursery, kindergarten, elementary, middle, high school and university* **) and list them, <b> one per box </b>, in the boxes below.**
*If yesterday, during your school day you had close contact with a considerable number of people, report only the most significant ones (interactions lasting more than 15 continuous minutes).*
*Note: if among these people there is one that you have already included previously, DO NOT include them in the list below.*

[possible contact 1]

[possible contact 2]

[possible contact 3]

[possible contact 4]

[possible contact 5]

… etc …

<966> No new contacts to report *MOVE ABOVE*

---

**[ c_homeguest ] Think of all the </b> people </b> you had <b> <u>close contact with YESTERDAY</u> at <u>OTHER PEOPLE'S HOUSES</u> </b> and list them, <b> one per box </b >, in the boxes below.**
*Note: If any of these people are people you have already listed, DO NOT list them in the following list.*

[possible contact 1]

[possible contact 2]

[possible contact 3]

[possible contact 4]

[possible contact 5]

… etc …

<966> No new contacts to report *MOVE ABOVE*

---

 **[ c_restaurant ] If <b> <u>YESTERDAY</u> </b> you went to any <b> <u>RESTAURANT PLACE</u> </b> (** *for example, bar, restaurant, ice cream parlor, pizzeria, pastry shop* **), think back to any <b> people <b> with whom you had <b> <u>close contact</u> </b> on these occasions (** *for example, with people sitting at the table/counter with you, with waiters or shop assistants* **) and list them, <b> one per box </b>, in the boxes below.**
*Note: if among these people there is one that you have already included previously, DO NOT include them in the list below.*

[possible contact 1]

[possible contact 2]

[possible contact 3]

[possible contact 4]

[possible contact 5]

   … etc …

<966> *No new contacts to report MOVE ABOVE*

---

 **[ c_leisure ] If <b> <u>YESTERDAY</u> </b> you did any <b> <u>SPORTS OR PLEASURE ACTIVITY</u> </b> (for example, you went to the cinema, to the gym/swimming pool, to the park, you volunteered, etc.) think back to the <b> people <b> with whom you had <b> <u>close contact</u> </b> on all these occasions and list them, <b> one per box </b>, in the boxes below.**
*Note: if among these people there is one that you have already included previously, DO NOT include them in the list below.*

[possible contact 1]

[possible contact 2]

[possible contact 3]

[possible contact 4]

[possible contact 5]

   … etc …

<966> *No new contacts to report MOVE ABOVE*

---

 **[ c_shopping ] If <b> <u>YESTERDAY</u> </b> you went <b> <u>SHOPPING</u> </b> (in shops, supermarkets, markets, shopping centres), think back to the <b> people <b> with whom you had <b> <u>close contact</u> </b> on these occasions (for example, with people who accompanied you, shop assistants or other customers) and list them, <b> one per box </b>, in the boxes below.**

*Note: if among these people there is one that you have already included previously, DO NOT include them in the list below.*

[possible contact 1]

[possible contact 2]

[possible contact 3]

[possible contact 4]

[possible contact 5]

    … etc …

<966> *No new contacts to report*

*MOVE UP*

---

**[ c_transport ] If** <b> **YESTERDAY** </b> **you used one or more** <b> **MEANS OF TRANSPORT** </b> **(including a car), think back to the** <b> **people** </b> **you had** <b> **close contact with** </b> **on those occasions and list them,** <b> **one per box** </b> **>, in the boxes below.**
***Note: if any of these people are those you have already included previously, DO NOT include them in the list below.***

[possible contact 1]

[possible contact 2]

[possible contact 3]

[possible contact 4]

[possible contact 5]

    … etc …

<966> *No new contacts to report*

*MOVE UP*

---

**[ c_otherindoor ] If** <b> **YESTERDAY** </b> **you had** <b> **close contact INDOOR occasions** other **than those presented so far** </b> **list the** <b> **people** </b> **with whom you have had them,** <b> **one per box** </b> **, in the boxes below.**
***Note: if among these people there are any that you have already entered previously, DO NOT enter them in the list below.***

[possible contact 1]

[possible contact 2]

[possible contact 3]

[possible contact 4]

[possible contact 5]

   … etc …

<966> *No new contacts to report MOVE ABOVE*

---

*#Base: Total*
*#Filter: if isolation != 1*
*#Receiverif not adult : child , if respondent_type_LF =1 ->*
   *if respondent_type = 2 insert blue bold warning message "This question concerns you, ["son" if*
   *childgender=1] ["daughter"if childgender=2] aged [$pipe-in:childage] as the recipient of this*
   *survey"*
*#Question type: open text boxes*

**[ c_otheroutdoor ] If <b> <u>YESTERDAY</u> </b> you had <b> <u>close contact OUTDOORS other than those presented so far</u> </b> , please list the <b> people <b> with whom you had them, <b> one per box </b>, in the boxes below.**
   ***Note: If any of these people are people you have already listed previously, DO NOT list them in the following list.***

[possible contact 1]

[possible contact 2]

[possible contact 3]

[possible contact 4]

[possible contact 5]

   … etc …

<966> *No new contacts to report MOVE ABOVE*

---

*#TOSCRIPTING :*

*PLEASE CREATE A **TEST MODE** VARIABLE WHERE THERE WILL BE THE TOTAL COUNT (SUM)*
   *OF CONTACTS WRITTEN IN ALL THE ABOVE MODULE 2B OPEN QUESTIONS*
   *(eg 3 open boxes filled in c_home + 2 filled in c_work = 5 in total)*

**[ total_contacts ]** *[TOTAL COUNT TO APPEAR] – please insert also c_coviventi rows=1 in the count*

---

*#TOSCRIPTING :*

*PLEASE IN **TEST MODE** LET ALL THE CONTACTS APPEAR (all the contacts filled in c_home to*
   *c_otheroutdoor questions)*

**[ contacts_name ]** *[total contact list to appear] please insert also c_cohabitants rows=1 in the count*

*#toscripting: <u>from closed to c_vaccino questions</u> , please also add pipe in names as they appear from c_conviventi rows=1*

---

*#Base: Had at least one close contact*

**[closed] Please indicate which of the following contacts occurred <u>mostly</u> indoors and which mostly outdoors. Select one response per line.**

 -**[closed_1** *]#TOSCRIPTING: insert* **Contact name 1** *had $pipein[PLACE]*

 -**[closed_2** *]#TOSCRIPTING: insert* **Contact name 2** *had $pipein[PLACE]*

 -**[closed_3** *]#TOSCRIPTING: insert* **Contact name 3** *had $pipein[PLACE]*

 -**[closed_4** *]#TOSCRIPTING: insert* **Contact name 4** *had $pipein[PLACE]*

 -**[closed_5]** *#TOSCRIPTING: insert* **Contact name 5** *had $pipein[PLACE]*

 *… AND SO ON depending on number of contacts in [total_contacts]*

**#TOSCRIPTING: *[PLACE]* TABLE:**

**Please note you've inverted c_home and c_homeguest pips in**

| *In which question was contact written* | *Text to write in rows $ pipein* |
|---|---|
| *C_cohabitants* | had at home |
| *C_home* | *had at home* |
| *C_school* | *had at school* |
| *C_work* | *had at work* |
| *C_transport* | *had on a means of transport* |
| *C_shopping* | *got while shopping* |
| *C_restaurant* | *had in a catering establishment* |
| *C_leisure* | *had in my spare time* |
| *C_homeguest* | *had at someone else's house* |
| *C_otherindoor* | *#as by email Mar 1, please do not show this but count it in total_contacts_indoor* |
| *C_outdoor* | *#as by email Mar 1, please do not show this* |

<1>Indoors
<2>Outdoor

---

*#TOSCRIPTING :*

*PLEASE CREATE A* **TEST MODE** *VARIABLE WHERE THERE WILL BE THE TOTAL COUNT (SUM) OF CONTACTS THAT HAD CODE 1 "AL INDOORS" IN [CLOSED] QUESTION*

**[ total_contacts_indoor ]** *[TOTAL COUNT TO APPEAR]*

---

*#Base: Total*
*#Filter*

**[ c_sharedindoor ] <u>HOW</u> MANY PEOPLE DID YOU <u>SHARE INDOOR SPACES WITH</u> <u>YESTERDAY</u> for an extended period of time <u>(</u> more than 30 minutes)?**

**Please also include the close contact people listed above if you shared an indoor space with them for more than 30 minutes.**
**For example, classmates if you are a student, or office/work colleagues if you are a worker, a compartment in a transport vehicle, etc.**

***Please provide your best estimate in the box below.***
*#toscripting: re-shape this with the spaces as above*

[INTEGER NUMBER]

*#TOSCRIPTING: in all these module questions please show contacts in rows (randomize order and keep same order on different questions).*
*freeze the TOP row of grid*

---

*#Base: Had at least one close contact*
*#Filter: [ total_contacts ]>=1*
*#Receiver if not adult: child*
*#Question type: open integer grid, ~~random rows~~*
   *#toscripting: please do not show cohabitants ( cohabitants ) here as we already know their age, cohabitants will be shown from c_fisico on*
**[ c_age ] How old are the following people? Write the age of each person in the appropriate box. *If you don't know the exact age, write the age you think they are.***

 -[c_eta_1]contact name 1

 -[c_eta_2]contact name 2

 -[c_eta_3]contact name 3

 -[c_eta_4]contact name 4

 -[c_eta_5]contact name 5

*... AND SO ON DEPENDING ON [total_contacts]*

[TYPE NUMBER] *(FROM 0 TO 100)*

---

*#Base: Had at least one close contact*
*#Filter: [ total_contacts ]>=1*
*#Receiverif not adult : child , if respondent_type LF =1 ->*
   *if respondent_type = 2 insert blue bold warning message "This question concerns you, ["son" if childgender=1] ["daughter"if childgender=2] aged [$pipe-in:childage] as the recipient of this survey"*
*#Question type: single grid, ~~random rows~~*
**[ c_fisico ] There was <u>physical contact of any kind</u> ~~(e.g. did you touch each other)~~ in yesterday's meeting with the following people? Select one answer per person.**

 -[c_fisico_1]contact name 1

 -[c_fisico_2]contact name 2

 -[c_fisico_3]contact name 3

 -[c_fisico_4]contact name 4

 -[c_fisico_5]contact name 5

*... AND SO ON DEPENDING ON [total_contacts]*

<1>Yes

<2>No

<977>~~I don't remember~~

---

*#Base: Had at least one close contact*
*#Filter: [ total_contacts ]>=1*
*#Receiverif not adult : child , if respondent_type LF =1 ->*
   *if respondent_type = 2 insert blue bold warning message "This question concerns you, ["son" if*

*#Question type: single grid, ~~random rows~~, random reverse order for columns*

**[ c_mask ] Which of you wore a <u>mask the entire time</u> ? Select one answer per person.**

 -[c_mask_1]contact name 1

 -[c_mask_2]contact name 2

 -[c_mask_3]contact name 3

 -[c_mask_4]contact name 4

 -[c_mask_5]contact name 5

*... AND SO ON DEPENDING ON [total_contacts]*

<1>Neither *fixed*

<2>Only me

<3>Only this person

<4>Both

<977 fixed>I don't remember

---

*#Base: Had at least one close contact*
*#Filter: [ total_contacts ]>=1*
*#Receiver if not adult: child, if respondent_type_LF =1 **AND childage** >=5 ->*
*#Question type: single grid, ~~random rows~~. ~~Required - SOFT~~*
   *if respondent_type = 2 insert blue bold warning message "This question concerns you, ["son" if childgender=1] ["daughter"if childgender=2] aged [$pipe-in:childage] as the recipient of this survey"*

**[ c_reddito_percepito ] In your opinion, on a scale of 1 to 5, what level of economic well-being do the following people rank in relation to the Italian population?**
   **Select one answer per person.**

 -[c_received_income_1]contact name 1

 -[c_received_income_2]contact name 2

 -[c_received_income_3]contact name 3

 -[c_received_income_4]contact name 4

 -[c_received_income_5]contact name 5

*... AND SO ON DEPENDING ON [total_contacts]*

<1>1 = very low well-being

<2>2 = low well-being

<3>3 = average well-being

<4>4 = high well-being

<5>5 = very high well-being

<977>I don't know

---

*#Base: Had at least one close contact*
*#Filter: [ total_contacts ]>=1*
*#Receiver if not adult: child*
*#Question type: single grid*

**[ c_istruzione ] What is the highest level of education achieved by the following people?**
   **Select one answer per person.**

 -[c_instruction_1]contact name 1 *$pipe-in [c_age] as follows, in every row:* contact name 1 ( *$pipe-in* years)

-[c_statement_2]contact name 2 *same $pipe-in in all rows*
-[c_istruzione_3]contact name 3
-[c_ instruction_4]contact name 4
-[c_istruzione_5]contact name 5

*... AND SO ON DEPENDING ON [total_contacts]*

<1>Middle school diploma (or less)

<2>High school diploma (lyceum, technical or vocational institute) *not selectable if [c_eta] (same row)<=17*

<3>Degree or higher (master's, doctorate, etc.) *not selectable if [c_eta] (same row)<=20*

<977>I don't know

#TOSCRIPTING: warning message if inconsistency in code 2 and code 3: "Some of the education levels listed are not plausible for the age of the person in question, please review your answers"
PLEASE ALLOW THE RESPONDENTS TO go ahead after showing warning message twice.

*#Base: Had at least one close contact*
*#Filter: [ total_contacts ]>=1*
*#Receiverif not adult : child , if respondent_type_LF =1 ->*
    *if respondent_type = 2 insert blue bold warning message "This question concerns you, ["son" if childgender=1] ["daughter"if childgender=2] aged [$pipe-in:childage] as the recipient of this survey"*
*#Question type: single grid, ~~random rows~~, Required=SOFT (leave it =SOFT, but adapt warning message " Please provide a response for at least one contact ", make the message appear ONLY IF RESPONDENT does not select any answer in any row)*

**[ c_relationship ] What is your relationship with the person contacted? Select one answer per person.**

-[c_relation_1]contact name 1
-[c_relation_2]contact name 2
-[c_relation_3]contact name 3
-[c_relationship_4]contact name 4
-[c_relation_5]contact name 5

*... AND SO ON DEPENDING ON [total_contacts]*
    *#toscripting: auto-assign code 1 for rows coming from c_coviventi in this question*

<1>Cohabiting relative/family member or roommate

<2>Non-cohabiting relative/family member

<3>Friend or acquaintance

<4>Co-worker *#show if respondent_type=3 AND [occupation]=1,2,3,6,7,8,9. or Show if respondent_type=2 AND childage respondent >=16 AND [occupation]=1,2,3 ,6,7,8,9*

<5>School/College Mate *#show if respondent_type=1 or (respondent_type=2 childage respndent=14,15) or (respondent_type=2,3 AND occupation=5,6,7*

<6>Other

*#Base: Had at least one close contact*
*#Filter: [ total_contacts ]>=1*
*#Receiverif not adult : child , if respondent_type_LF =1 ->*
    *if respondent_type = 2 insert blue bold warning message "This question concerns you, ["son" if childgender=1] ["daughter"if childgender=2] aged [$pipe-in:childage] as the recipient of this survey"*

**[ c_frequency ] On how many days have you had <u>close contact</u> with this person in the past 14 days (including today)? Select one answer per person.**

***If you don't remember the exact number, select your best estimate.***

 -[c_frequency_1]contact name 1

 -[c_frequency_2]contact name 2

 -[c_frequency_3]contact name 3

 -[c_frequency_4]contact name 4

 -[c_frequency_5]contact name 5

*… AND SO ON DEPENDING ON [total_contacts]*

<1>1

<2>2

<3>3

<4>4

<5>5

<6>6

<7>7

<8>8

<9>9

<10>    10

<11>    11

<12>    12

<13>    13

<14>    14

**[ c_gender ] What is the gender of the following people? Select one answer per person.**

 -[c_gender_1]contact name 1

 -[c_gender_2]contact name 2

 -[c_gender_3]contact name 3

 -[c_gender_4]contact name 4

 -[c_gender_5]contact name 5

 *… AND SO ON DEPENDING ON [total_contacts]*

<1>Male

<2>Female

**[ c_distance ] What minimum distance was maintained during the following non-physical contacts? Select one answer per person. *If you do not remember, select the "I do not remember" option or simply do not answer for that person.***

-[c_distance_1]contact name 1

-[c_distance_2]contact name 2

-[c_distance_3]contact name 3

-[c_distance_4]contact name 4

-[c_distance_5]contact name 5

*... AND SO ON DEPENDING ON [total_contacts]*

<1>Less than 1 meter

<2>Between 1 meter and 2 meters

<3>More than 2 meters

<977 fixed>I don't remember

**[ c_vaccino ] Have the following people already received one or more doses of the COVID-19 vaccine? Select one answer per person. *If you do not remember, select the option "I don't know/I don't remember." ~~or simply do not answer for that person.~~***

-[c_vaccino_1]contact name 1

-[c_vaccino_2]contact name 2

-[c_vaccino_3]contact name 3

-[c_vaccino_4]contact name 4

-[c_vaccino_5]contact name 5

*... AND SO ON DEPENDING ON [total_contacts]*

<1>Yes

<2>No

<977>I don't know/I don't remember

**[child013_presence] Were you able to get the support of [pipe-in: "your son" if childgender=1, "your daughter" if childgender=2] of $pipein** *[childage]* **years in answering the questions just asked about contacts made yesterday?**
***If [pipe-in: "your son" if childgender=1, "your daughter" if childgender=2] was not* [pipe-in: "could" if childgender=1, "could" if childgender=2] *present, this is not a problem* .**

<1>Yes, he was present
<2>No, he wasn't present

*#TOSCRIPTING: show this module only to adults AND children whose childhood>=14*

*#Base: Total ( respondent (aged 14+)*
*#Filter: age or child age >=14*
*#Receiverif not adult : child ( if child age >=14), if childage <=13: parent*
*if respondent_type = 2 insert blue bold warning message "This question concerns you, ["son" if childgender=1] ["daughter"if childgender=2] aged [$pipe-in:childage] as the recipient of this survey"*
*#Question type: single grid, randomize rows (same order of pers_risk rows) – skip option health*

**[inf] <u>Do you personally know at least one person</u> who has contracted, been hospitalized, or died from each of the following diseases? Select one answer per disease.**

-[inf_sarscov] COVID-19

-[inf_flu]Influenza

-[inf_measles]Measles

<1>I only know someone who has contracted the disease

<2>I also know someone who has been hospitalized (but no one who has died)

<3>I also know someone who died

<4>*I don't know anyone who has contracted the disease* # toscripting : make italic

*#Base: Total (5+)*
*#Filter: if age or childage >=5*
*#Receiver if not adult: child, if respondent_type_LF =1 ->*
  *if respondent_type = 2 insert blue bold warning message "This question concerns you, ["son" if childgender=1] ["daughter"if childgender=2] aged [$pipe-in:childage] as the recipient of this survey"*
*#Question type: single - skip option health*

**[ vacc_covid ] What is your SARS-CoV-2 (COVID-19) vaccination status?**
  **Select one answer only.**

<1>I am not vaccinated by choice
<2>    I am not vaccinated because I have a health exemption
<3>    I have already booked the appointment for the first dose
<4>    I got the first dose, but not the second
<5>    I did the second dose, but not the booster (third dose)
<6>    I did the booster but not the fourth dose
<7>    I did the booster and the fourth dose
<955fixed>Other (specify) *[SPECIFY]*

*#Base: Total (5+)*
*#Filter: if age or childage >=5, if vacc_covid =1*
*#Receiver if not adult: PARENT, if respondent_type_LF =1OR 2 ->*
*#Question type: multiple, randomize - skip option health*

**[ no_vax_reason ] What motivated your choice not to get vaccinated?**
  **Select all that apply.**

<1>I have doubts about the effectiveness of the vaccine
<2>I have doubts about the (health) safety of the vaccine
<3>I am against mandatory vaccination
<4>I am against it because covid-19 does not pose a serious threat to my health.
<5>I am against the green pass
<6>I was advised against it by an acquaintance
<7>I haven't had time to inform myself properly yet
<8>I think the vaccine is a waste of money for the benefit of big pharmaceutical companies
<955fixed>Other (specify) *[SPECIFY]*

*#Base: Total*
*#Filter:*
*#Receiverif not adult : child , if respondent_type_LF =1 ->*
  *if respondent_type = 2 insert blue bold warning message "This question concerns you, ["son" if childgender=1] ["daughter"if childgender=2] aged [$pipe-in:childage] as the recipient of this survey"*
*#Question type: single - skip option health*

**[ vacc_altri ] Are you <u>vaccinated against at least one</u> of the following diseases: Human Papilloma Virus (HPV), Meningococcus, Pneumococcus, Hepatitis A, Herpes Zoster/Shingles?**
**Select one answer only.**

<1>Yes

<2>No

<977>I don't know/I don't remember

---

**[refuse] Have you ever actively refused a vaccine when it was offered to you?**
**Select one answer only.**

<1>Yes

<2>No

<977>I don't know/I don't remember

---

**[ txt_final ] You have completed the questionnaire, we thank you very much for your contribution!**